\documentclass[fleqn,usenatbib,onecolumn]{mnras}

\usepackage[T1]{fontenc}

\DeclareRobustCommand{\VAN}[3]{#2}
\let\VANthebibliography\thebibliography
\def\thebibliography{\DeclareRobustCommand{\VAN}[3]{##3}\VANthebibliography}

\usepackage{graphicx}
\usepackage{amsmath}	
\usepackage{amssymb}	
\usepackage{bm}
\usepackage{hyperref}
\usepackage[normalem]{ulem}

\usepackage{newtxtext,newtxmath} 

\let\vec\bm
\newcommand{\uvec}[1]{\vec{\hat{#1}}}

\newcommand{\diff}{\ensuremath{\mathrm{d}}}
\newcommand{\e}{\mathrm{e}}
\newcommand{\I}{\mathrm{i}}
\newcommand{\ints}{\!\!\int\!\!}
\newcommand{\sdot}{\!\cdot\!}

\newcommand{\erf}{\mathrm{erf}}
\newcommand{\erfc}{\mathrm{erfc}}
\newcommand{\Ei}{\mathrm{Ei}}

\newcommand{\Si}{\ensuremath{\mathrm{Si}}}
\newcommand{\Ci}{\ensuremath{\mathrm{Ci}}}

\newcommand{\dnx}{\diff^3\vec x}
\newcommand{\dnu}{\diff^3\vec u}
\newcommand{\dnq}{\frac{\diff^3\vec q}{(2\pi)^3}}
\newcommand{\dnqprime}{\frac{\diff^3\vec q^\prime}{(2\pi)^3}}
\newcommand{\dqq}{\frac{\diff q}{q}}

\newcommand{\ddelta}{\delta_D}
\newcommand{\ddeltan}{\delta_D^3}
\newcommand{\step}{\theta_H}

\newcommand{\bmax}{b_\mathrm{max}}
\newcommand{\bmin}{b_\mathrm{min}}

\newcommand{\V}{V}
\newcommand{\vV}{\vec V}

\newcommand{\vVi}{\vec{V}_\mathrm{imp}}

\newcommand{\deltas}{\delta_*}
\newcommand{\Ps}{P_*}

\newcommand{\Dv}{\Delta\vec v}

\newcommand{\rhob}{\bar\rho}

\def\vx{\vec{x}}
\def\vv{\vec{v}}
\def\vu{\vec{u}}
\def\vq{\vec{q}}
\def\vr{\vec{r}}
\def\uvx{\uvec{x}}
\def\uvv{\uvec{v}}
\def\uvu{\uvec{u}}
\def\uvq{\uvec{q}}
\def\uvr{\uvec{r}}

\newcommand{\vorb}{v_\mathrm{orb}}
\newcommand{\vvorb}{\vv_\mathrm{orb}}
\newcommand{\uvvorb}{\uvv_\mathrm{orb}}

\def\ba#1\ea{\begin{align}#1\end{align}}
\def\refeq#1{Eq.~(\ref{#1})}

\title[Stellar streams and substructure]{Stellar streams and dark substructure: the diffusion regime}

\author[M. S. Delos \& F. Schmidt]{
M. Sten Delos,$^{1}$\thanks{E-mail: sten@mpa-garching.mpg.de (MSD); fabians@mpa-garching.mpg.de (FS)}
Fabian Schmidt$^{1}$\footnotemark[1]
\\
$^{1}$Max Planck Institute for Astrophysics, Karl-Schwarzschild-Str. 1, 85748 Garching, Germany
}

\date{Accepted XXX. Received YYY; in original form ZZZ}

\pubyear{2021}

\begin{document}
\label{firstpage}
\pagerange{\pageref{firstpage}--\pageref{lastpage}}
\maketitle

\begin{abstract}
  The cold dark matter picture predicts an abundance of substructure within the Galactic halo. However, most substructures host no stars and can only be detected indirectly. Stellar streams present a promising probe of this dark substructure. These streams arise from tidally stripped star clusters or dwarf galaxies, and their low dynamical temperature and negligible self-gravity give them a sharp memory of gravitational perturbations caused by passing dark substructures. For this reason, perturbed stellar streams have been the subject of substantial study. While previous studies have been largely numerical, we show here that in the diffusion regime -- where stream stars are subjected to many small velocity kicks -- stream perturbations can be understood on a fully analytic level. In particular, we derive how the (three-dimensional) power spectrum of the substructure density field determines the power spectrum of the (one-dimensional) density of a stellar stream. Our analytic description supplies a clear picture of the behaviour of stream perturbations in response to a perturbing environment, which may include contributions from both dark and luminous substructure. In particular, stream perturbations grow in amplitude initially, settle into a steady state, and ultimately decay. By directly relating stellar stream perturbations to the surrounding matter distribution, this analytic framework represents a versatile new tool for probing the nature of dark matter through astrophysical observations.
\end{abstract}




\section{Introduction}

Stellar streams arise from the gradual tidal stripping of member stars from globular clusters or other objects in the gravitational potential of the Milky Way. These stars roughly follow the orbit of the parent cluster, or progenitor, but due to their velocity dispersion they spread out to form an approximately one-dimensional extension \citep[e.g.][]{eyre/binney:2011}. Stellar streams are interesting probes of dark matter because their negligible self-gravity, coupled with the low velocity dispersion of their member stars (of the order of 10~km\,s$^{-1}$), allows them to retain a sharp memory of past gravitational perturbations. Perturbations to stellar streams therefore reflect the presence and properties of any substructure in the Galactic dark matter halo \citep{ibata/etal:2002,2002ApJ...570..656J,2008ApJ...681...40S}, an abundance of which is predicted in most cold dark matter scenarios \citep[e.g.][]{10.1111/j.1365-2966.2008.14066.x}.

Searches for substructure-induced stream perturbations initially focused on single encounters with fairly massive subhaloes, which can lead to gaps in the streams or other major features \citep{carlberg:2009,2011ApJ...731...58Y,sanders/etal:2016,bonaca/etal:2019,2021MNRAS.501..179M,2021A&A...647A.137J,2021ApJ...911..149L,2022AJ....163...18F,2022ApJ...925..118T}. Such encounters are rare, however, and clear gaps only form for sufficiently disruptive encounters. Hence, recent literature has begun to consider observables which probe the effects of more numerous encounters with smaller substructure. For example, \citet{banik/etal:2018,2021MNRAS.502.2364B,banik/etal:2019} used the one-dimensional power spectrum of the stellar density along a stream to place constraints on warm dark matter, a model of dark matter that predicts less substructure. These constraints were based on the forward model in action-angle space presented in \citet{bovy2017linear}. \citet{bovy2017linear} studied encounters between a stream and $\mathcal{O}(100)$ subhaloes within the mass range $(10^5,10^9)~\mathrm{M}_\odot$, while \citet{banik/etal:2018} modelled about 25 encounters within the mass range $(10^6,10^9)~\mathrm{M}_\odot$.

Our goal in this paper is to further illuminate the regime of many small encounters. In particular, we derive a direct, analytic connection between the power spectrum $\mathcal{P}(q)$ of the substructure density field and the resulting power spectrum $\Ps(k)$ of the (one-dimensional) stellar density along a stream. Note that we use $q$ to indicate the three-dimensional wavenumber within the environment and reserve $k$ for the one-dimensional wavenumber on the stream. Apart from elucidating the relevant physics that connects substructure encounters with the perturbed stream density in a statistical sense, our results allow for a simple and direct computation of the phenomenological effect of various substructure components, dark or luminous, on stellar streams. Ultimately, we hope that this approach can be used to place model-independent, purely gravitational constraints on the amplitude of substructure on different mass scales in the Milky Way halo.

The approach developed here rests on the approximation of many small velocity kicks acting on a given star in the stream. In this regime, \emph{stream stars diffuse}, i.e. perform a random walk, around their expected trajectory in the smooth halo potential. Also, while we allow the perturbing substructure to be described by a general density field, we make simplifying assumptions about the substructure velocity distribution; for instance we neglect the impact of internal dynamics within subhaloes. Finally, we employ approximate descriptions of orbital dynamics and global stream evolution that, in principle, only become valid when the number of orbital periods is large. We find, however, that only a couple of orbital periods are indeed necessary before our analytic predictions match numerical simulation results. Ultimately, despite these approximations, the match between analytic predictions and simulation results turns out to be remarkably tight.

The outline of the paper is as follows. In Section~\ref{sec:formalism}, we develop a procedure by which the statistics of integrated velocity injections due to an inhomogeneous environment may be derived from the density power spectrum $\mathcal{P}(q)$ of that environment. In Section~\ref{sec:inhomogeneity}, we use this procedure to express the one-dimensional power spectrum $P_{\Delta v}(k)$ of velocity injections into a stellar stream as a function of $\mathcal{P}(q)$. In Section~\ref{sec:disp}, we treat the stream's response to these velocity injections, ultimately expressing the stream density power spectrum $\Ps(k)$ as a function of $P_{\Delta v}(k)$. We also validate this treatment on idealized simulations.
Section~\ref{sec:orbit} presents our approximate treatment of orbital dynamics, while Section~\ref{sec:stream} presents our approximate treatment of a stellar stream's global (unperturbed) evolution. In Section~\ref{sec:disc}, we put our treatment of stellar stream perturbations into practice: we compare analytic predictions to full-stream simulation results, and we discuss some of the analytic framework's many implications. We conclude in Section~\ref{sec:conclusion}. Finally, Appendices \ref{sec:subfield_demo}--\ref{sec:unequaltime}, \ref{sec:bmax}, and~\ref{sec:sigmalocal} present mathematical arguments left out of the body of the article; Appendix~\ref{sec:pkv} presents a straightforward extension to this work by deriving the power spectrum $P_v(k)$ of stellar velocities within a stream; and Appendix~\ref{sec:simulation} details the idealized Monte Carlo simulations that we perform to validate our analytic derivations in Section~\ref{sec:disp}.

\section{Velocity injection formalism}\label{sec:formalism}

A stellar stream's self-gravity is negligible, so each star can be considered independently. The star resides within an inhomogeneous environment composed of dark matter substructure, which perturbs the star's motion through its gravity. In this section we develop the relationship between the statistics of the substructure environment and the integrated velocity change $\Delta \vv$ that it injects on to the star.\footnote{We refer to these $\Delta\vv$ as velocity injections.}

\subsection{Static substructure environment}\label{sec:static}

We begin with a simplified picture. Consider a star inside a substructure environment with static density field $\rho(\vx)=\rhob[1+\delta(\vx)]$, where $\rhob$ is the average density. If the star is at position $\vr$, it experiences acceleration
\begin{equation}
    \dot{\vec v}(\vr) = G\rhob\int\dnx\delta(\vx)\frac{\vx-\vr}{|\vx-\vr|^3},
\end{equation}
where the constant term $\rhob$ in the density has cancelled by symmetry. Substituting the inverse Fourier transform $\delta(\vq)\equiv (2\pi)^{-3}\int\diff^3\vq\,\e^{\I\vq\cdot\vx}\delta(\vq)$ and carrying out the $\vx$ integral, we find that\footnote{As shorthand, for any vector $\vx$ we write $x\equiv |\vx|$ and $\uvx\equiv \vx/x$. Recall, also, that we use $\vq$ to denote the three-dimensional wavenumber within the environment and reserve $k$ for the one-dimensional wavenumber on the stream, which we will introduce in Section~\ref{sec:inhomogeneity}.}
\begin{equation}\label{acceleration}
    \dot{\vec v}(\vr) = -4\pi \I G\rhob\int\dnq\delta(\vq)\e^{\I\vq\cdot\vr}\frac{\vq}{q^2}.
\end{equation}
Since we are interested in perturbations to the stream, \refeq{acceleration} does not include the acceleration due to the large-scale gravitational potential that governs the entire stream orbit. We will discuss in Section~\ref{sec:orbit} the impact of orbital dynamics on our treatment.

\begin{figure}
	\centering
	\includegraphics[width=\textwidth,trim=0cm 8.7cm 0cm 2.2cm,clip=true]{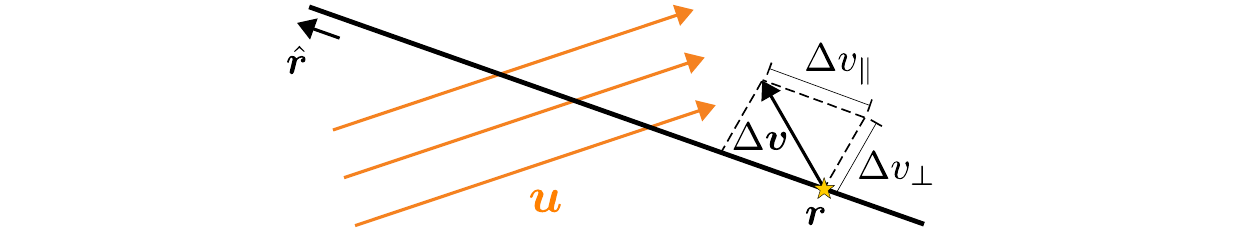}
	\caption{Sketch of the set-up studied here. The thick black line indicates the stellar stream, oriented along the direction $\uvr$, while the orange arrows mark the relative velocity of the perturbing substructure environment. A star located at position $\vr$ on the stream experiences a velocity kick $\Delta\vv$, which has components $\Delta v_\parallel$ along the stream and $\Delta\bm{v}_\perp$ perpendicular to the stream.}
	\label{fig:sketch}
\end{figure}

Now suppose there is a relative velocity $\vu$ between the star's (unperturbed) orbital velocity and the substructure environment so that in relation to the environment, the star's position at time $t$ is $\vr-\vu t$ (see Fig.~\ref{fig:sketch}).\footnote{In general $\vv$ will denote the velocity of a star while $\vu$ will denote a velocity associated with the substructure environment, although at this point in our treatment only changes to $\vv$ are meaningful.} Integrated over the duration $t$, the change in the star's velocity is
\begin{align}
    \Dv(\vr) &= -4\pi \I G\rhob\int\dnq\delta(\vq)\e^{\I\vq\cdot\vr}\frac{\vq}{q^2}\int_0^t\diff t^\prime \e^{-\I\vq\cdot\vu t^\prime}
    \\\label{Dv_k}
    &=
    \int\dnq\delta(\vq)\e^{\I\vq\cdot\vr}\vV^*(\vq|\vu,t),
\end{align}
where we assume $\Delta v \ll u$ so that the relative trajectory is not significantly perturbed by the velocity injection.\footnote{In later sections, we take $\Dv(\vr)$ to represent the velocity injection \textit{on the point $\vr$}, rather than on a particular star. Therefore, our neglect here of the star's response does not affect later computations.} We also assume that the substructure environment is not perturbed by the presence of the star. In Eq.~(\ref{Dv_k}), we define
\begin{equation}\label{V}
    \vV(\vq|\vu,t)\equiv 8\pi \I G\rhob \,\e^{\I\vq\cdot\vu t/2}\frac{\sin(\vq\cdot\vu\, t/2)}{\vq\cdot\vu}\frac{\vq}{q^2}.
\end{equation}
This very useful function has dimensions of velocity and represents the response of $\Dv$ (over the time interval $t$) to substructure at the wavenumber $\vq$ moving at velocity $\vu$. We will also define $\vV(\vx|\vu,t)$ as the inverse Fourier transform of $\vV(\vq|\vu,t)$, so that
\begin{equation}\label{Dv_x}
    \Dv(\vr) = \int\dnx\, \delta(\vx)\vV(\vx-\vr|\vu,t).
\end{equation}
We will not need to evaluate $\vV(\vx|\vu,t)$ explicitly.

\subsection{General substructure velocity distribution}\label{sec:general}

We can straightforwardly apply the results of Section~\ref{sec:static} to the scenario where the substructure environment has a velocity distribution $f(\vu)\diff^3\vu$. To do so, let us decompose the density contrast field $\delta(\vx)$ into $N$ components $\delta_i(\vx)$ such that
\begin{align}\label{delta_decomp}
    \delta(\vx)&=\sum_{i=1}^N\delta_i(\vx).
\end{align}
We take each subfield $\delta_i(\vx)$ to move with velocity $\vu_i$ sampled from the distribution $f(\vu)$. In the continuum limit of this velocity distribution, the summation is therefore replaced by the integral
\begin{equation}\label{sum-to-v-integral}
    \frac{1}{N}\sum_{i=1}^N \to \int\diff^3\vu f(\vu).
\end{equation}
The results of Section~\ref{sec:static} may now be applied to each subfield individually using the decomposition in Eq.~(\ref{delta_decomp}) and the result integrated over the velocity distribution using Eq.~(\ref{sum-to-v-integral}). We illustrate this process in Appendix~\ref{sec:subfield_demo}.

We aim to study how correlations between velocity injections depend on the power spectrum $P(q)$ of the substructure density field. For this purpose, we assume additionally that all of the density subfields $\delta_i(\vx)$ are independent and have the same power spectrum; that is (in Fourier space)
\begin{equation}\label{delta_independence}
    \langle\delta_i(\vq)\delta_j^*(\vq^\prime)\rangle
    = 
    (2\pi)^3\delta_{ij} \ddeltan(\vq-\vq^\prime) P_N(q),
\end{equation}
where $\delta_{ij}$ is the Kronecker delta, $\ddeltan$ is the three-dimensional Dirac delta function, and $P_N(q)$ is the power spectrum associated with each of the $N$ density subfields $\delta_i(\vx)$. It follows that
\begin{equation}\label{subpower}
    P_N(q)=P(q)/N,
\end{equation}
where $P(q)$ is the power spectrum associated with the full density field $\delta(\vx)$. Compared to a completely arbitrary substructure phase-space distribution, we have made two simplifying assumptions: that the density power spectrum of substructure particles moving at velocity $\vu$ is independent of $\vu$, and that the density fields associated with particles moving at different velocities $\vu$ are independent. Roughly, these simplifications mean that we neglect internal motion within subhaloes.\footnote{The assumption that particles moving at different velocities have the same power spectrum is similar to, but weaker than, the assumption that the distribution function is separable, i.e. $f(\vx,\vu)=\rho(\vx)f_\mathrm{vel}(\vu)$. A separable distribution function (with a non-trivial velocity distribution) cannot support persistent inhomogeneity, however.}

\subsection{Example: the mean squared velocity injection}\label{sec:v2}

\begin{figure}
	\centering
	\includegraphics[width=0.5\linewidth]{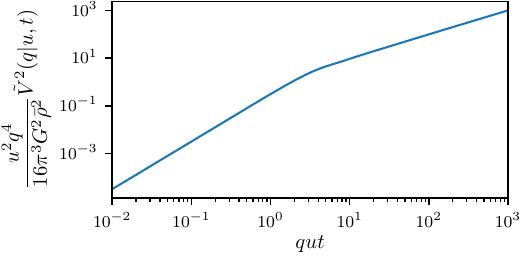}
	\caption{Response function $\tilde \V^2(q|u,t)$ for the mean squared velocity injection $\langle\Dv^2\rangle$ to substructure at wavenumber $q$ moving at velocity $u$. The function transitions from $\tilde \V^2\propto t^2$ when $qut\ll 1$ (corresponding to a sustained coherent acceleration) to $\tilde \V^2\propto t$ when $qut\gg 1$ (corresponding to a random walk).}
	\label{fig:V}
\end{figure}

We now demonstrate a few important aspects of the formalism with an example calculation. We first show in Appendix~\ref{sec:subfield_demo} that the subfield construction in Section~\ref{sec:general} implies that the correlation between two velocity injections is
\begin{align}
	\label{vcorr_full}
	\langle\Dv(0)\cdot\Dv(\vec r)\rangle
	&=
	\int\!\!\dnq P(q)\e^{-\I \vq \cdot\vr}\int\diff^3\vu f(\vu) |\vV(\vq|\vu,t)|^2.
\end{align}
If we specialize to the case $\vec r=0$ and carry out the angular integrals over $\vq$ in Eq.~(\ref{vcorr_full}), we find that
\begin{align}\label{mean_dv2}
	\langle\Dv^2\rangle
	&=
	\int_0^\infty\dqq\mathcal{P}(q) \int\diff^3\vu f(\vu) \tilde \V^2 (q|u,t),
\end{align}
where $\mathcal{P}(q)\equiv [q^3/(2\pi^2)]P(q)$ is the dimensionless power spectrum and $\tilde \V^2(q|u,t)$ is the spherical average (over $\vq$) of $|\vV(\vq|\vu,t)|^2$. In particular,
\begin{align}\label{V_tilde}
	\tilde \V^2(q|u,t) &= 64\pi^2\frac{G^2\rhob^2}{u^2 q^4}\int_{-1}^{1}\frac{\diff\mu}{2}  \frac{\sin^2(\mu qut/2)}{\mu^2},
\end{align}
where the integration variable is $\mu\equiv \uvq\cdot\uvu$. The function $\tilde \V^2(q|u,t)$ has dimensions of squared velocity and represents the contribution to the variance $\langle\Dv^2\rangle$ that arises from substructure power at the wavenumber $q$ and velocity $u$. We plot $\tilde \V^2(q|u,t)$ in Fig.~\ref{fig:V}.

While $\tilde \V^2$ can be evaluated analytically (in terms of the sine integral function), it is more insightful to instead examine two limiting cases:\footnote{The dichotomy between short and long time-scales is also discussed, using a different analytic formulation, in \citet{10.1093/mnras/stz2648}.}
\begin{enumerate}
	\item If $qut < 1$, the relative distance traversed between the star and substructure is smaller than the scale of inhomogeneity in the substructure environment. In this case, $\langle \Dv^2\rangle\propto\tilde \V^2\propto t^2$ because velocity kicks add coherently. Mathematically, in this limit $\sin(\mu qut/2)\simeq \mu qut/2$, so $ \tilde \V^2(q|u,t) \simeq 16\pi^2 G^2\rhob^2 q^{-2} t^2$.
	\item If $qut \gg 1$, the relative distance traversed is much larger than the scale of substructure inhomogeneity. In this case, $\langle \Dv^2\rangle\propto t$ because the velocity injection undergoes a random walk.
\end{enumerate}
The latter case $qut \gg 1$ is the most relevant scenario observationally because the age $t$ of any stellar system is typically far longer than the substructure encounter time-scale $(qu)^{-1}$. For instance, if $u\simeq 200~\mathrm{km}\,\mathrm{s}^{-1}$ and $q\simeq 1~\mathrm{kpc}^{-1}$, then $(qu)^{-1}\simeq 5$~Myr, far shorter than the multiple-Gyr lifetimes of stellar streams. The clearest way to see the behaviour of $\tilde \V$ in the $qut \gg 1$ regime is to take advantage of the delta-function identity
\begin{equation}\label{sin2}
    \lim_{a\to\infty}\frac{\sin^2(a \mu)}{a \mu^2}=\pi \ddelta(\mu),
\end{equation}
where $\ddelta$ is the Dirac delta function. Setting $a= q u t/2$, it follows from Eq.~(\ref{V_tilde}) that
\begin{equation}\label{Vlim}
    \tilde \V^2(q|u,t) = \frac{16\pi^3 G^2 \rhob^2}{u q^3} t,\ \  \text{ if } qut\gg 1 .
\end{equation}
Evidently, the $qut\gg 1$ limit enforces $\vq\cdot\vu=0$, i.e. that only modes $\vq$ lying in the plane perpendicular to $\vu$ contribute to integrated velocity injections.

It follows from Eqs. (\ref{mean_dv2}) and~(\ref{Vlim}) that in the $qut\gg 1$ limit, the mean velocity dispersion $\langle\Dv^2\rangle$ induced by the substructure environment is
\begin{equation}\label{mean_sigma2}
	\langle\Dv^2\rangle = 16\pi^3 G^2\rhob^2\left\langle\frac{1}{u}\right\rangle t \int_0^\infty\dqq\mathcal{P}(q) q^{-3}.
\end{equation}
Here, $\langle 1/u\rangle$ represents the average of $1/u$ over the velocity distribution $f(\vu)$. Note the dependence on the low-$u$ tail of the distribution, which is naturally expected as smaller relative velocities lead to larger kicks. We also remark that if the substructure environment has a characteristic mass scale $M_0 \propto R_0^3$, then the integral over the power spectrum $P(q)$ and hence $\langle\Dv^2\rangle$ scale as $M_0$ (see also Section~\ref{sec:halo}). In fact, Eq.~(\ref{mean_sigma2}) is consistent with Eq.~(2.5) in \cite{2018arXiv180800464A}, who adopted a particle-based treatment of the heating of stellar streams, if we associate $R_0^3 = r_{\rm sc}^3$, $\Delta v \simeq \omega_z v_c$ and $u\simeq v_c$. We will later explore the connection between density-field-based and particle-based substructure treatments more precisely using the correspondence discussed in Section~\ref{sec:halo} between the density field power spectrum and the particle mass spectrum.

We can also evaluate the angular integrals over $\vq$ in Eq.~(\ref{vcorr_full}) in the $qut\gg 1$ limit for general $\vec r$. In this case we find that
\begin{align}
    \langle\Dv(0)\cdot\Dv(\vec r)\rangle
    =
    16\pi^3 G^2\rhob^2 t
    \int_0^\infty\dqq\mathcal{P}(q) q^{-3}\int\diff^3\vu f(\vu) u^{-1} J_0(q r_\perp),
    \label{vcorr}
\end{align}
where $J_0$ is a Bessel function of the first kind and $r_\perp\equiv\sqrt{\vr^2-(\vr\cdot\uvu)^2}$ is the component of $\vr$ perpendicular to $\vu$. To interpret this result, we note that Eq.~(\ref{V}) implies the velocity injection contributed by substructure at the wavenumber $\vq$ points along $\vq$. Since $qut\gg 1$ enforces $\vq\cdot\vu=0$, the velocity injection lives in the two-dimensional plane perpendicular to $\vu$. For each value of $\vu$, we can thus think of $\mathcal{P}(q) \tilde \V^2 (q|u,t)=16\pi^3 G^2\rhob^2 t\, \mathcal{P}(q) q^{-3} u^{-1}$ as (roughly) the dimensionless power spectrum of two-dimensional velocity perturbations, whose correlation function is given by an integral over $\dqq$ with the Bessel function $J_0(q r_\perp)$, where $r_\perp$ is the lag in the plane perpendicular to $\vu$. We treat correlated velocity injections and their power spectra more thoroughly in the next section.

\section{Correlated velocity injections}\label{sec:inhomogeneity}

We now use the formalism of Section~\ref{sec:formalism} to more precisely derive how velocity injections at different points on a stellar stream are correlated.

\subsection{The velocity-injection power spectrum}

A procedure similar to that in Section~\ref{sec:v2} yields the velocity-injection correlation function
\begin{align}\label{corr_dv}
	\xi_{\uvec a\uvec b}(\vr)\equiv \langle \uvec a\sdot\Dv(0) \uvec b\sdot\Dv(\vr)\rangle
	&=
	\ints\dnq P(q)\ints\diff^3\vu f(\vu)
	\e^{-\I\vq\cdot\vec r}
	\uvec a\sdot\vV^*(\vq|\vu,t)\uvec b\sdot\vV(\vq|\vu,t)
\end{align}
along arbitrary unit vectors $\uvec a$ and $\uvec b$. As in Section~\ref{sec:v2}, the $qut\gg 1$ limit enforces $\uvec q\cdot\uvec u=0$ through a delta function:
\begin{align}\label{corr_dv_}
	\xi_{\uvec a\uvec b}(\vr) &= 32\pi^3 G^2\rhob^2 t
    \ints\dnq\frac{P(q)}{q^3}
    \ints\dnu \frac{f(\vu)}{u}
    \e^{-\I\vq\cdot\vr}
    (\uvec a\cdot\uvq)(\uvec b\cdot\uvq)
    \ddelta(\uvq\cdot\uvu).
\end{align}
The remaining derivation in this section will continue to assume the $qut\gg 1$ limit.

The angular integrals over $\vq$ in Eq.~(\ref{corr_dv_}) can be evaluated explicitly in terms of Bessel functions (similarly to Eq.~\ref{vcorr}). However, it is convenient to instead Fourier transform Eq.~(\ref{corr_dv_}) to obtain the one-dimensional power spectrum
\begin{align}\label{pow_dv}
	P_{\uvec a\uvec b}(k) 
	&\equiv \int_{-\infty}^\infty \diff r\, \e^{-\I k r} \xi_{\uvec a\uvec b}(\vr)
	= 32\pi^3 G^2\rhob^2 t
    \ints\dnq\frac{P(q)}{q^3}
    \ints\dnu \frac{f(\vu)}{u}
    2\pi \ddelta(\vq\cdot\uvr+k)
    (\uvec a\cdot\uvq)(\uvec b\cdot\uvq)
    \ddelta(\uvq\cdot\uvu).
\end{align}
It is notable that due to the delta function $\ddelta(\vq\cdot\uvr+k)$, which arises from the integral over $r$, correlated velocity injections at the scale $k$ arise only due to substructure modes $\vq$ whose projection on to the stream is equal to $k$, i.e. $|\vq\cdot\uvr|=k$.

Now let us assume that anisotropy in the substructure environment's relative velocity distribution $f(\vu)$ arises entirely due to the stellar stream's own orbital motion, which we take to be parallel to $\uvr$. That is, we assume that the substructure has an isotropic velocity distribution in the Galactic frame. In Appendix~\ref{sec:angular}, we evaluate the angular $\vq$ integrals in Eq.~(\ref{pow_dv}) and show that under this assumption,\footnote{To be precise, Eq.~(\ref{Pvpara1}) holds for arbitrary $f(\vu)$, while Eq.~(\ref{Pvperp1}) relies on the assumption that $f(\vu)$ depends on the direction of $\vu$ only through $\uvu\cdot\uvr$ (i.e. the angle between $\vu$ and the stream). This assumption is weaker than the assumption that the substructure velocity distribution is isotropic in the Galactic frame.} the only two independent velocity-injection power spectra are 
\begin{align}\label{Pvpara1}
    P_{\Delta v,\parallel}(k) &\equiv P_{\uvr\uvr}(k) 
    =
    32\pi^3 G^2\rhob^2 k^2 t
    \int_k^\infty\dqq\frac{\mathcal{P}(q)}{q^6}
    \ints\dnu \frac{f(\vu)}{u}
    \frac{\step[1-(\uvu\cdot\uvr)^2-k^2/q^2]}{[1-(\uvu\cdot\uvr)^2-k^2/q^2]^{1/2}}
    \\\label{Pvperp1}
    P_{\Delta v,\perp}(k) &\equiv P_{\uvec p\uvec p}(k) 
    =
    16\pi^3 G^2\rhob^2 t
    \int_k^\infty\dqq\frac{\mathcal{P}(q)}{q^6}
    \left(q^2-k^2\right)
    \ints\dnu \frac{f(\vu)}{u}
    \frac{\step[1-(\uvu\cdot\uvr)^2-k^2/q^2]}{[1-(\uvu\cdot\uvr)^2-k^2/q^2]^{1/2}},
\end{align}
where $\uvec p$ is any unit vector perpendicular to the stream (so $\uvec p\cdot\uvr=0$) and $\mathcal{P}(q)\equiv[q^3/(2\pi^2)]P(q)$ again. Here, the Heaviside unit step functions $\step$ ensure that only substructure wavenumbers $q$ larger than $k$ contribute, and we changed the integration limits to emphasize this point. There are no velocity-injection correlations on scales smaller than those at which the substructure environment is inhomogeneous.\footnote{Equation~(\ref{pow_dv}) implied that the condition $|\vec q\cdot\uvr|=k$ was required for the substructure mode $\vq$ to induce correlated velocity injections on the stream at the wavenumber $k$. The integration of $\vec q$ over angles has turned this equality condition into the inequality $q>k$.}

We can simplify these expressions further. Let $\tilde\vu$ be the substructure velocity in the Galactic frame and $\tilde f(\tilde u)$ be its isotropic velocity distribution (which depends only on $\tilde u\equiv|\tilde\vu|$). Additionally, let $v$ be the magnitude of the stream's velocity in the Galactic frame, which is parallel to $\uvr$. We show in Appendix~\ref{sec:anisotropy} that
\begin{align}\label{Pvpara}
    P_{\Delta v,\parallel}(k) &=
    16\pi^4 G^2\rhob^2 k^2 t
    \int_k^\infty\dqq\frac{\mathcal{P}(q)}{q^6}
    \int\diff^3\tilde\vu \frac{\tilde f(\tilde u)}{\tilde u}
    \step\!\left(\tilde u-\frac{k}{q}v\right)
    &\text{(isotropic substructure velocities)}
    \\\label{Pvperp}
    P_{\Delta v,\perp}(k) &=
    8\pi^4 G^2\rhob^2 t
    \int_k^\infty\dqq\frac{\mathcal{P}(q)}{q^6}
    \left(q^2-k^2\right)
    \int\diff^3\tilde\vu \frac{\tilde f(\tilde u)}{\tilde u}
    \step\!\left(\tilde u-\frac{k}{q}v\right).
    &\text{(isotropic substructure velocities)}
\end{align}
Beyond the $q>k$ enforced by the integration limits, the step functions $\step$ here also enforce $q\tilde u>kv$; that is, the time-scale $(q\tilde u)^{-1}$ associated with substructure passage must be shorter than the time-scale $(kv)^{-1}$ associated with the passage of the subject stream at the given wavenumber $k$.\footnote{By combining the explicit step function in Eqs. (\ref{Pvpara}) and~(\ref{Pvperp}) with the integration limit, we may notice that the step function of Eqs. (\ref{Pvpara1}) and~(\ref{Pvperp1}) has become $\step(q-\max\{1,v/\tilde u\}k)$. One interpretation of the role of $v$ therein is as follows. Recall that the $qut\gg 1$ limit enforces $\vq\perp\vu$. Raising the stream velocity $v$ realigns the relative substructure velocity $\vu=\tilde\vu-\vv$ to be closer, in angle, to $-\vv$, which points along the stream. Thus, $\vq$ becomes closer to perpendicular to the stream, which implies that a larger $q=|\vq|$ is needed to achieve $|\vq\cdot\uvr|=k$ and thereby inject velocities at the scale $k$ (see Eq.~\ref{pow_dv}).} It is interesting to note that the stream's velocity $v$ has no impact outside of this step function. For instance, in no sense is $v$ additive with the substructure velocity $\tilde\vu$ except inside the step function. Intuitively this finding may be explained by the idea that motion of substructure parallel to the stellar stream perturbs the whole stream coherently (at least in the $qut\gg 1$ limit) and therefore does not induce inhomogeneity.

To simplify these expressions still further we must assume a form for the Galactic-frame substructure velocity distribution $\tilde f(\tilde u)$. Let us assume that $\tilde f(\tilde u)$ is Maxwellian with velocity dispersion $u_0$ per dimension; that is,
\begin{equation}
    \tilde f(\tilde u)\diff^3\tilde \vu =
    \sqrt{\frac{2}{\pi}}\frac{\tilde u^2\diff\tilde u}{u_0^3}
    \exp\left(-\frac{\tilde u^2}{2u_0^2}\right).
\end{equation}
In this case the velocity integral in Eqs. (\ref{Pvpara}) and~(\ref{Pvperp}) yields the factor $\langle 1/\tilde u\rangle=\sqrt{2/\pi}/u_0$ with the step function yielding an additional exponential factor:
\begin{align}\label{Pvpara_MB}
    P_{\Delta v,\parallel}(k) &=
    16\pi^4 G^2\rhob^2\frac{\sqrt{2/\pi}}{u_0} k^2 t
    \int_k^\infty\dqq\frac{\mathcal{P}(q)}{q^6}
    \exp\!\left(-\frac{1}{2}\frac{k^2}{q^2}\frac{v^2}{u_0^2}\right)
    &\text{(Maxwellian substructure velocities)}
\end{align}
and likewise for $P_{\Delta v,\perp}(k)$. Typically one might expect the stream velocity $v$ to be of order $u_0$, in which case the exponential factor has only a modest impact because $k<q$ is already enforced. Larger stream velocities $v$, however, can lead to very strongly suppressed power spectra.

\subsection{Connection to subhalo models}\label{sec:halo}

Our treatment of velocity injections above is expressed in terms of the substructure power spectrum $P(k)$ (or its dimensionless variant $\mathcal{P}$).
In theoretical treatments of dark matter substructure, it is common to predict instead the subhalo mass function $\diff n/\diff M$, which specifies the differential number density $n$ of subhaloes per subhalo mass $M$ \citep[e.g.][]{gao2011statistics}. Each halo's internal structure, in turn, is described by a density profile that is often taken to be universal up to rescalings in density and radius \citep[e.g.][]{navarro1997universal}, or equivalently, mass and radius.
Given these ingredients, it is straightforward to compute the power spectrum.

Suppose a halo of mass $M$ and characteristic scale radius $R$ has the density profile
\begin{align}
	\rho(r|M,R) &= M R^{-3} p(r/R),
\end{align}
where $p(x)$, the universal density profile, is a dimensionless function of a dimensionless argument. If halo positions are uncorrelated and the halo radius $R$ is a function of mass $M$, then the substructure power spectrum is
\begin{equation}
  \label{Pksub}
	P(k) = \frac{1}{\rhob^2} \int \diff M \frac{\diff n}{\diff M} \left[M\tilde p(k R)\right]^2
\end{equation}
\citep[]{scherrer1991statistics}, where $\rhob$ is the mean density of the substructure environment\footnote{$\rhob$ is arbitrary as long as the same value is used in the expressions in Sections \ref{sec:formalism} and~\ref{sec:inhomogeneity}. So, for instance, we could choose to also include any smooth background density in the definition of $\rhob$.} and
\begin{equation}
    \tilde p(x) \equiv \int_0^\infty 4\pi y^2\diff y p(y) \frac{\sin(x y)}{x y}
\end{equation}
is the three-dimensional Fourier transform of $p(x)$. For two simple density profiles,
\begin{align}\label{plummer}
    p(x)&=[3/(4\pi)] (1+x^2)^{-5/2}, 
    &&\tilde p(x) = x K_1(x),
    && \text{\citep{plummer1911problem}}
    \\\label{hernquist}
    p(x)&=[1/(2\pi)] x^{-1} (1+x)^{-3}, 
    &&\tilde p(x)=1-(\pi/2) x\cos x - x\,\Ci(x)\sin x + x\,\Si(x)\cos x,
    && \text{\citep{hernquist1990analytical}}
\end{align}
where $K_1$ is a modified (hyperbolic) Bessel function of the second kind, $\Si(z)=\int_0^z \diff t \sin (t)/t$ is the sine integral function, and $\Ci(z)=-\int_z^\infty \diff t \cos (t)/t$ is the cosine integral function. The assumption that $R$ is a function of $M$ is a common simplification \citep[e.g.][]{diemer2019accurate}, but Equation~(\ref{Pksub}) may be straightforwardly extended to also integrate over the scatter in $R$ at fixed mass. We also comment that Eqs. (\ref{Pvpara_MB}) and~(\ref{Pksub}) may be combined to write the power spectrum of velocity injections as an integral,
\begin{align}
    P_{\Delta v,\parallel}(k) &=
    8\pi^2 G^2\frac{\sqrt{2/\pi}}{u_0} k^2 t
    \int \diff M \frac{\diff n}{\diff M} M^2
    \int_k^\infty\!\!\diff q \frac{\left[\tilde p(q R)\right]^2}{q^4}
    \exp\!\left(-\frac{1}{2}\frac{k^2}{q^2}\frac{v^2}{u_0^2}\right),
    &\text{(subhaloes with Maxwellian velocities)}
\end{align}
over the subhalo mass spectrum.

While this procedure shows how to convert a halo model into a power spectrum, it is important to note that the power spectrum is a more generally applicable description. Inhomogeneity may exist within the stream's dark matter environment that cannot be represented as a collection of subhaloes; these subhaloes' own tidal streams represent one example. A more exotic example is given by the quantum interference effects associated with fuzzy or ultralight dark matter \citep[e.g.][]{2014NatPh..10..496S,10.1093/mnras/staa738,10.1093/mnras/stab1764,2021JCAP...03..076D}.
In this scenario, the density is given by the square $|\psi|^2$ of a complex wavefunction $\psi$, leading to interference patterns and hence significant density variations on the scale of the de Broglie wavelength $\hbar/(m_{\rm FDM} u)$ (for particle velocity $u$). These density fluctuations are ubiquitous but probably not well approximated as static subhaloes. Hence, the approach laid out in this paper, which does not rely on a subhalo prescription but only requires the specification of the power spectrum $P(k)$ of the total substructure density (e.g. measured in simulations), is well suited to investigating the phenomenological effects of fuzzy dark matter. This is the subject of ongoing work.

\section{Perturbations to the stream}\label{sec:disp}

In Section~\ref{sec:inhomogeneity}, we derived the one-dimensional power spectrum of integrated velocity injections into a stellar stream. In this section we now derive the stream's response in phase space to these velocity injections. Since we focus on the regime of many small encounters, it is appropriate to treat the effect of the encounters as a diffusive process via the Fokker--Planck approach. That is, the displacement of an individual star by a single encounter can be assumed to be much less than the width of the phase-space distribution function $f(x,v,t)$ of the stream stars. Further, we assume that spatially varying perturbations to the distribution function remain small, so that we can work to linear order in perturbations around a uniform zeroth-order distribution function $f_0(v,t)$. Note, however, that we still allow $f_0(v,t)$ to be altered by the continuous heating induced by encounters.

Let the stream's distribution function thus be $f(x,v,t)=f_0(v,t)+f_1(x,v,t)$, where $f_0$ is spatially uniform and $f_1$ is a small spatially varying perturbation that we will treat at linear order. In the Fokker--Planck approximation, $f$ satisfies the Boltzmann equation
\begin{equation}\label{BE}
    \frac{\partial f}{\partial t} + v \frac{\partial f}{\partial x} = - C\frac{\partial f}{\partial v} + \frac{1}{2} D \frac{\partial^2 f}{\partial v^2},
\end{equation}
where we assume that the diffusion coefficients $C=(\diff/\diff t)\Delta v$ and $D=(\diff/\diff t)\Delta v^2$, which represent the velocity injection and its square per unit time, respectively, depend on position $x$ but not velocity $v$. These diffusion coefficients connect directly to the results of Section~\ref{sec:inhomogeneity}. In particular, $C$ is the external acceleration of stream stars along the stream track and satisfies
\begin{align}\label{Ccorr}
    \left\langle C(k,t) C^*(k^\prime,t^\prime)\right\rangle
    =
    2\pi\ddelta(k-k^\prime)\ddelta(t-t^\prime)\frac{\diff P_{\Delta v,\parallel}(k,t)}{\diff t}
\end{align}
if we assume that velocity kicks at different times $t$ and $t^\prime$ are uncorrelated. This assumption amounts to assuming that substructure encounters are impulsive, and we also show explicitly in Appendix~\ref{sec:unequaltime} that it holds in the long-time $qut\gg 1$ and $qut^\prime\gg 1$ limit (see Section~\ref{sec:v2}). Note that $C$ is a perturbative quantity in that its spatial average is 0. Meanwhile, the second coefficient $D$ is in principle the integrated velocity-injection power per unit time,
\begin{align}\label{Dinf}
  D = \int_{-\infty}^\infty\frac{\diff k}{2\pi} \frac{\diff P_{\Delta v,\parallel}(k,t)}{\diff t},
\end{align}
although in practice it can be somewhat more complicated (see Section~\ref{sec:D}). We will neglect any spatial and temporal variation in $D$, but we discuss the impact of this approximation in Section~\ref{sec:D}.

Separating Eq.~(\ref{BE}) into equations for the spatial average and the perturbations yields
\begin{align}\label{BE0}
    \frac{\partial f_0}{\partial t} &= \frac{1}{2} D \frac{\partial^2 f_0}{\partial v^2},
    \\\label{BE1}
    \frac{\partial f_1}{\partial t} + v \frac{\partial f_1}{\partial x}
    &=
    -C(x,t)\frac{\partial f_0}{\partial v}
    +\frac{1}{2} D \frac{\partial^2 f_1}{\partial v^2},
\end{align}
where we neglect terms at second order or higher in the perturbations. Taking the second velocity moment of Eq.~(\ref{BE0}), we find that the stream's spatially averaged velocity dispersion $\sigma^2(t)$ grows as
\begin{align}\label{sigma2}
    \sigma^2(t) = \sigma_0^2 + D t,
\end{align}
where $\sigma_0^2$ is the stream's initial velocity dispersion at the time $t=0$. The perturbation equation is more challenging, but we solve it in Appendix~\ref{sec:f1} and show that if the unperturbed distribution function $f_0$ is Maxwellian with (one-dimensional) velocity dispersion given by Eq.~(\ref{sigma2}), then the perturbation to the distribution function is (in Fourier space)
\begin{align}\label{f1}
    f_1(k,v,t) &= \rhob_* \int_{0}^t\diff t^\prime
    C(k,t^\prime)
    \frac{v-\I k D(t-t^\prime)^2/2}{\sqrt{2\pi}(\sigma_0^2+D t)^{3/2}}
    \exp\!\left[-\I kv(t-t^\prime)-\frac{1}{6} k^2 D (t-t^\prime)^3
    -\frac{(v-\I k D(t-t^\prime)^2/2)^2}{2(\sigma_0^2+D t)}\right],
\end{align}
where $\rhob_*$ is the stream's unperturbed density.

\subsection{Stream density perturbations}\label{sec:density}

We can immediately integrate Eq.~(\ref{f1}) over velocities and divide by $\rhob_*$ to obtain the stream density contrast
\begin{align}\label{delta-vdisp}
    \deltas(k,t) &= -\I k 
    \int_{0}^t\diff t^\prime
    (t-t^\prime) C(k,t^\prime)
    \e^{-k^2 (t-t^\prime)^2 [\sigma_0^2+D(t+2t^\prime)/3]/2}.
\end{align}
We remark that if the velocity dispersion $\sigma_0^2+Dt$ can be neglected, then $\deltas(k,t) = - \I k s(k,t)$, where $s(k,t)=\int_0^t\diff t^\prime \Delta v(k,t^\prime)$ represents the displacement of stars in response to velocity injections.
This is the result that an effective fluid treatment would predict: that the density contrast is the negative divergence of the displacement field.
It follows from Eq.~\eqref{delta-vdisp} when we note that $\Delta v(k,t)=\int_0^t\diff t^\prime C(k,t^\prime)$ and integrate by parts to obtain $s(k,t)={\int_0^t\diff t^\prime\, (t-t^\prime) C(k,t^\prime)}$.

\subsubsection{Density perturbation response to solitary velocity kicks}\label{sec:densitypert}

\begin{figure*}
	\centering
	\includegraphics[width=\linewidth]{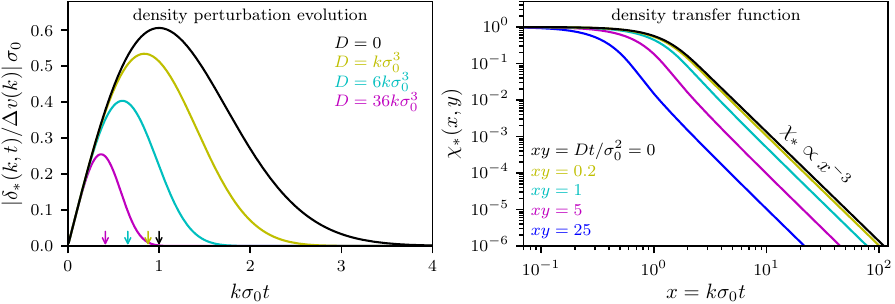}
	\caption{Behaviour of stream density perturbations. \textit{Left-hand panel}: Time evolution of the stream's density contrast $\delta_*$ due to some distribution $\Delta v(k)$ of velocity kicks that occur at time $t=0$. The system's velocity dispersion is $\sigma_0^2+Dt$ and suppresses any density or velocity perturbation after a brief growth period. Different colours represent systems with different levels of the velocity dispersion growth rate $D$. In each case, we also indicate with a small arrow the time when the rms particle displacement (due to the velocity dispersion) is precisely $k^{-1}$; that is, $k s_\mathrm{rms}=1$ (see Eq.~\ref{srms}). \textit{Right-hand panel}: The transfer function $\chi_*$ that sets the power spectrum $\Ps(k,t)$ of stream density according to Eq.~(\ref{powerd}). The black curve shows the case where the initial velocity dispersion $\sigma_0$ dominates over the velocity dispersion induced by the substructure environment, while the coloured curves show scenarios with increasingly significant induced velocity dispersions. Note that by fixing the combination $xy$ for each curve, we fix the time $t$, so that $x\propto k$ and this plot shows how the power spectrum is scaled as a function of $k$.}
	\label{fig:cutoff-d}
\end{figure*}

As a simple case, let us explore the stream's response to velocity kicks that are restricted to one moment in time, which we take to be $t=0$. In this scenario $C(k,t')=\Delta v(k)\ddelta(t')$, and the velocity injection $\Delta v(k,t)=\Delta v(k)$ is constant in time (after $t=0$). We still allow the velocity dispersion to grow in time at the rate $D$ (so if $D>0$, then we are studying the response to any one of a continuous series of velocity kicks). It now follows from Eq.~(\ref{delta-vdisp}) that the stream density contrast evolves as
\begin{align}\label{delta_time}
    \delta_*(k,t)
    &= -\I k t\,\e^{-k^2 (\sigma_0^2+D t/3)t^2/2} \Delta v(k)
    &&
    \text{(constant $\Delta v$)}
\end{align}
We plot the resulting time evolution of $\delta_*$ in the left-hand panel of Fig.~\ref{fig:cutoff-d}. The initial velocity kick $\Delta v$ causes $\delta_*$ to grow proportionally with time before being exponentially suppressed by the velocity dispersion. Indeed, one can show (e.g. using random walk theory) that if the velocity dispersion grows as $\sigma^2=\sigma_0^2+Dt$ due to random kicks, then the rms particle displacement arising therefrom is given by
\begin{align}\label{srms}
s_\mathrm{rms}\equiv(\sigma_0^2 + D t/3)^{1/2} t
\end{align}
The square of this expression appears in the exponent in Eq.~(\ref{delta_time}). Thus, we expect that the exponential suppression makes $\delta_*$ negligible if $k s_\mathrm{rms}\gtrsim 3$. This notion is borne out in Fig.~\ref{fig:cutoff-d}, where we indicate where $ks_\mathrm{rms}=1$ using a small arrow.

\subsubsection{Density power spectrum from continuous velocity kicks}\label{sec:densitypower}

We now consider the stream's full response to ongoing velocity kicks due to the environment, and we write the stream density power spectrum $P_*(k,t)$ in terms of the power spectrum $P_{\Delta v,\parallel}$ of $\Delta v$. Using the definition $\langle\delta_*(k,t)\delta_*^*(k^\prime,t)\rangle\equiv 2\pi\ddelta(k-k^\prime)P_*(k)$, Eqs. (\ref{delta-vdisp}) and~(\ref{Ccorr}) imply
\begin{align}\label{powerd-time}
    \Ps(k,t) = k^2 
    \int_0^t\diff t^\prime (t-t^\prime)^2
    \exp\!\left[-k^2 \left(\sigma_0^2+D \frac{t+2t^\prime}{3}\right)(t-t^\prime)^2\right]
    \frac{\diff P_{\Delta v,\parallel}(k,t^\prime)}{\diff t^\prime}.
\end{align}
If we now specialize to the case where $P_{\Delta v,\parallel}\propto t$, as in Section~\ref{sec:inhomogeneity}, then $\diff P_{\Delta v,\parallel}(k,t^\prime)/\diff t^\prime = P_{\Delta v,\parallel}(k,t^\prime)/t^\prime$ is constant and we may pull it out of the integral. Then we can write the relationship
\begin{equation}\label{powerd}
    \Ps(k,t) = 
    \chi_*\!\left(k\sigma_0 t,\frac{D}{k\sigma_0^3}\right)
    \frac{k^2 t^2}{3}
    P_{\Delta v,\parallel}(k,t)
\end{equation}
between the density and velocity-injection power spectra in terms of a transfer function
\begin{align}\label{cutoffd}
\chi_*(x,y)\equiv
    \frac{3}{x^3}\int_0^x\diff x^\prime (x-x^\prime)^2 \exp\!\left[-\left(x-x^\prime\right)^2\left(1+y\frac{x+2x^\prime}{3}\right)\right]
\end{align}
that encodes the impact of the velocity dispersion $\sigma_0^2+Dt$. For convenience, we set the pre-factors in Eq.~(\ref{powerd}) such that $\chi_*=1$ in the limit that the velocity dispersion can be neglected.

The expression for the transfer function $\chi_*(x,y)$ must in general be integrated numerically, and we plot the behaviour of $\chi_*(x,y)$ in the right-hand panel of Fig.~\ref{fig:cutoff-d}. However, if $\sigma_0^2\gg D t$, we can approximate $y=0$ and evaluate
\begin{align}
    \chi_*(x,0) &=
    \frac{3\sqrt{\pi}}{4}x^{-3}\erf(x) - \frac{3}{2}x^{-2}\e^{-x^2},
\end{align}
where $\erf(x)\equiv (2/\sqrt{\pi})\int_0^x \diff t\,\e^{-t^2}$ is the error function. In this case we see that when $x=k\sigma_0 t\ll 1$, $\chi_*(x,0)\simeq 1$, and when $x\gg 1$, $\chi_*(x,0)\simeq (3/4)\pi^{1/2} x^{-3}$. At any fixed time $t$, this effect may be viewed as a free-streaming cut-off: at large scales $k\ll(\sigma_0 t)^{-1}$, the velocity dispersion has no significant effect, but at small scales $k\gg(\sigma_0 t)^{-1}$, the velocity dispersion-induced streaming of stars suppresses power by a factor proportional to $k^{-3}$.

We may also interpret these limits through the lens of time evolution. Since $\Ps\propto t^3\chi_*$, the same scalings imply that at any fixed $k$, the power spectrum initially grows as $\Ps\propto t^3$ (since $\chi_*\simeq 1$ here); specifically
\begin{align}\label{earlypower}
    \Ps(k,t) = \frac{k^2 t^3}{3} \frac{\diff P_{\Delta v,\parallel}(k,t)}{\diff t},
    &&
    \text{(growing regime)}
\end{align}
where we replace $P_{\Delta v,\parallel}/t$ with $\diff P_{\Delta v,\parallel}/\diff t$ to more clearly indicate the time evolution behavior. The growing phase ends when $t\sim (k\sigma_0)^{-1}$, after which $\Ps$ becomes constant (since $\chi_*\propto t^{-3}$ in this regime). When $t\gtrsim (k\sigma_0)^{-1}$, the system may be viewed as entering a steady state where the continuous injection of new power cancels the suppression of power by the velocity dispersion. The power spectrum is
\begin{align}\label{staticpower}
    \Ps(k,t) = 
    \frac{\pi^{1/2}}{4}
    k^{-1}\sigma_0^{-3} \frac{\diff P_{\Delta v,\parallel}(k,t)}{\diff t}
    &&
    \text{(steady-state regime)}
\end{align}
during this phase. Finally, we can show that when the induced velocity dispersion becomes non-negligible ($Dt\gtrsim\sigma_0^2$), the steady-state balance tips and the net density power begins to decrease over time. We can obtain the power in this regime by simply replacing $\sigma_0^2$ in Eq.~(\ref{staticpower}) with $Dt$, whence
\begin{align}\label{decaypower}
    \Ps(k,t) = 
    \frac{\pi^{1/2}}{4}
    k^{-1} D^{-3/2} t^{-3/2} \frac{\diff P_{\Delta v,\parallel}(k,t)}{\diff t},
    &&
    \text{(decaying regime)}
\end{align}
i.e. $\Ps\propto t^{-3/2}$. To see why this substitution suffices, note that if $k^2Dt^3\gg 1$ (a consequence of $k\sigma_0 t\gg 1$ and $Dt\gg\sigma_0^2$), then the integral in Eq.~(\ref{powerd-time}) is dominated by the contribution when $t^\prime\simeq t$. Thus we can replace the factor $t+2t^\prime$ in the exponent with $3t$, in which case $\sigma_0^2$ and $Dt$ are evidently interchangeable.

\subsection[Approximating the diffusion coefficient D]{Approximating the diffusion coefficient \textit{D}}\label{sec:D}

We approximated above that $D=(\diff/\diff t)\Delta v^2$ is spatially uniform, but in practice the spatial variation of $D$ cannot be neglected. Velocity injections can be strongly correlated over any given scale (according to Eq.~\ref{Pvpara}), and large-scale velocity injections induce only coherent motion -- rather than a velocity dispersion -- at smaller scales. As an extreme example, a velocity injection at a scale much larger than the stream's total length perturbs the entire stream coherently, so it does not induce any velocity dispersion. From this example it is clear that the spatial variation in $D$ cannot be treated perturbatively either. The perturbation in $D$ can be of order $D$ itself.

We have not found a solution to the Boltzmann equation, Eq.~(\ref{BE}), in the case where $D$ is a non-perturbative spatially varying quantity. In that solution's absence, we now discuss a suitable approximate expression for $D$. Recall that $D$ is essentially the integrated velocity-injection power per unit time; see Eq.~(\ref{Dinf}). To exclude the contributions of velocity injections that occur over scales larger than some maximum scale $1/k_\mathrm{min}$, a simple approach is to restrict the integration range so that
\begin{align}\label{D}
    D(k_\mathrm{min}) = 2\int_{k_\mathrm{min}}^\infty\frac{\diff k}{2\pi} \frac{\diff P_{\Delta v,\parallel}(k,t)}{\diff t}
\end{align}
(where the factor of 2 accounts for negative wavenumbers).

We first remark that if $Dt\ll \sigma_0^2$, then the choice of $D$ is immaterial as it does not impact predictions. We find in Section~\ref{sec:compare} that (at least for the substructure scenario we consider) setting $D=D(2\pi/L)$, where $L$ is the length of the stream, yields predictions that match simulation results reasonably well even when $Dt$ is moderately larger than $\sigma_0^2$. This definition allows velocity injections up to the scale of the stream to contribute to the velocity dispersion within the stream. When $Dt\gg \sigma_0^2$, this definition tends to overpredict the velocity dispersion at small scales and therefore underpredict the amount of stream power (since the velocity dispersion suppresses power). Hence, we may view predictions of $\Ps(k)$ with $D=D(2\pi/L)$ and $D=0$ as bracketing the true solution.

\subsection{Other stream statistics}\label{sec:other}

In Section~\ref{sec:density}, we derived the stream density power spectrum in terms of the power spectrum of velocity injections. It is straightforward to derive other properties of the stream in a similar way. For instance, in Appendix~\ref{sec:pkv} we derive the power spectrum $P_v$ of the mean stellar velocity within the stream. One can also express higher order density or velocity statistics, like the bispectrum, in terms of the corresponding statistics of velocity injections. We remark that in general, Eq.~(\ref{Dv_x}) implies that non-Gaussianity in velocity injections follows from non-Gaussianity in the (non-linear) substructure density-contrast field $\delta(\vx)$. Despite these possibilities, we will continue to focus on the power spectrum of stream density and leave consideration of other stream statistics to future work.

\subsection{Monte Carlo validation}\label{sec:validation}

\begin{figure*}
	\centering
	\includegraphics[width=\linewidth]{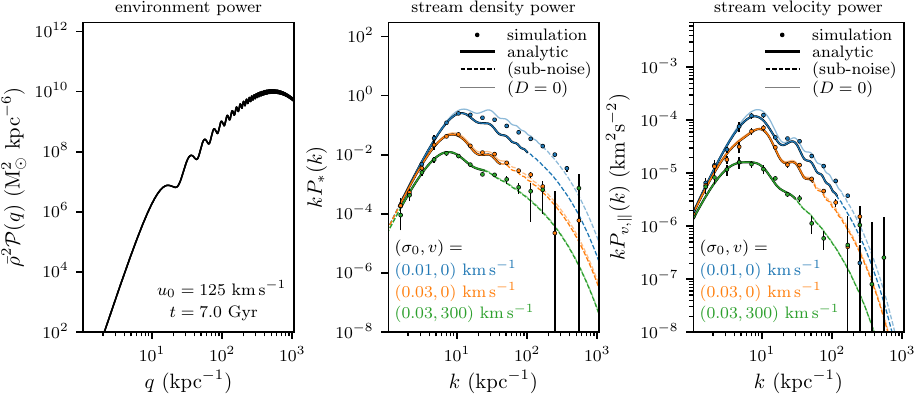}
	\caption{Validation of the computations in Sections \ref{sec:inhomogeneity} and~\ref{sec:disp} using idealized particle simulations. These simulations bombard a stream, represented as periodic line of stars $L=12$~kpc long, with a spectrum of extended ``subhalo'' particles that corresponds to the density power spectrum $\rhob^2\mathcal{P}(q)$ shown in the left-hand panel. Note that we impose a maximum impact parameter for encounters, which induces the oscillatory behavior (see Appendix~\ref{sec:bmax}). We also indicate the Maxwellian scale velocity $u_0$ and the simulation duration $t$. In the centre and right-hand panels, we compare the resulting stream density and velocity power spectra, respectively, to analytic predictions. We consider three different streams with different velocity dispersions $\sigma_0$ and average (orbital) velocities $v$; these are indicated by the different colours. Points indicate the mean result of 9 simulations, while error bars denote the standard deviation of the mean. The solid curves indicate the corresponding analytic predictions; these transition into dashed curves when they pass below the Poisson noise floor. Faint curves indicate the analytic predictions with zero induced velocity dispersion ($D=0$); see the discussion in Section~\ref{sec:D}. Above the Poisson noise floor (and even below it, to some extent), we find that our analytic predictions with $D=D(2\pi/L)$ (Eq.~\ref{D}) and those with $D=0$ successfully bracket the simulation results. Note that we subtract the noise floor from the simulation power spectra.}
	\label{fig:sim}
\end{figure*}

To verify the accuracy of our calculations, we carried out idealized simulations in which a periodic stellar stream is randomly subjected to encounters with passing extended ``subhalo'' particles, which we take to be Plummer spheres (Eq.~\ref{plummer}). The full construction of these simulations is detailed in Appendix~\ref{sec:simulation}, but we show the results in Fig.~\ref{fig:sim}. The left-hand panel shows the substructure environment's power spectrum $\mathcal{P}(q)$ and other properties. We consider three different stream models (denoted by different colours), and we plot both the density power spectrum (Section~\ref{sec:density}; centre panel) and the velocity power spectrum (Appendix~\ref{sec:pkv}; right-hand panel) of the stream. We find in all cases that our analytic predictions with $D=0$ (zero induced velocity dispersion) and $D=D(2\pi/L)$ (see Section~\ref{sec:D}) successfully bracket the simulation results, at least above the stream's Poisson noise floor. Moreover such bracketing is only relevant when the initial velocity dispersion is particularly small (blue scenario); typically the induced velocity dispersion $Dt$ does not become dominant. We explore this and other behaviours in Section~\ref{sec:disc} in the context of more realistic stream scenarios.

\section{Impact of orbital dynamics}\label{sec:orbit}

So far, we have viewed a stellar stream as an unbound one-dimensional system, but a real stellar stream orbits within the Galactic potential. A particular complication is that orbital dynamics make the connection between velocity kicks $\Delta \vv$ and the resulting position perturbations $\delta \vx$ non-trivial.\footnote{Orbital dynamics are trivial in action-angle space, and action-angle variables have consequently been employed in numerical treatments of stellar stream perturbations \citep[e.g.][]{bovy2017linear}. However, encounters with Galactic substructure occur in position-velocity space, so it is necessary to transform each velocity kick into action-angle space. Because of the extent to which this transformation complicates the derivation in Section~\ref{sec:inhomogeneity}, we do not employ action-angle coordinates in this study.} This is especially true given that stellar streams are often on highly eccentric orbits. In the present analysis we will take a simplified approach, considering the evolution of $\delta \vx$ and $\delta \vv$ over many orbits and regarding their behaviour over each orbit as a random variable. More precisely, for each velocity kick $\Dv$ we discard any information about when it occurred and instead assume it occurred at a uniformly distributed random time. We leave to future work a more precise accounting of orbital dynamics.

\subsection{Evolution of orbit perturbations}\label{sec:orbevo}

\begin{figure*}
	\centering
	\includegraphics[width=\linewidth]{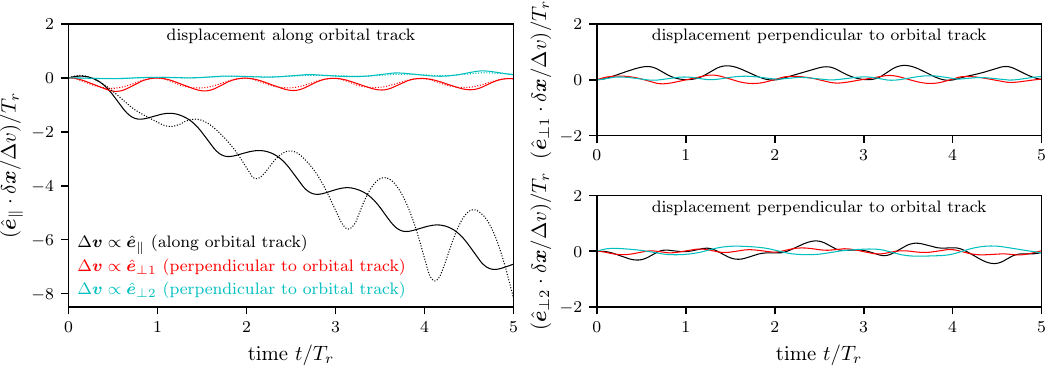}
	\caption{Evolution of the position perturbations parallel (left-hand panel) and perpendicular (right-hand panels) to the orbital velocity $\vvorb$, that result from an initial velocity kick $\Delta \vv$ at time $t=0$. We consider three directions of the velocity kick: one parallel to $\vvorb$ (black curves) and two perpendicular to $\vvorb$ (red and cyan). In particular, at each time $\uvec e_\parallel$, $\uvec e_{\perp 1}$, and $\uvec e_{\perp 2}$ form an orthonormal basis with $\uvec e_\parallel$ parallel to $\vvorb$. Only the parallel kick induces secular motion, and this secular motion only occurs along $\vvorb$ (black curves in the left-hand panel). In the left-hand panel, the dotted curves indicate the raw separation $(\vx-\vx_0)\cdot\uvvorb$ between perturbed and unperturbed orbits, while the solid curves indicate the separation with the overall expansion and contraction factored out (Eq.~\ref{pertframe}). Within the orbital context, $t=0$ occurs slightly after pericentre, when the radius is roughly $(3r_\mathrm{p}+r_\mathrm{a})/4$ with $r_\mathrm{p}$ and $r_\mathrm{a}$ the peri- and apocentre radii, respectively. $T_r$ is the radial orbit period.}
	\label{fig:dx}
\end{figure*}

To picture how perturbations evolve, we use \textsc{galpy} \citep[]{bovy2015galpy} to integrate particle orbits within a model of the Galaxy and its halo. For concreteness, we pick as the unperturbed orbit the orbit of the GD-1 stream progenitor as described in \citet{webb2019searching}. Figure~\ref{fig:dx} shows how the position perturbation $\delta\vx$ evolves due to a small velocity kick $\Delta \vv$. Within a coordinate system that is aligned along the orbital velocity vector $\vvorb$, a velocity kick generically excites oscillatory motion both parallel and perpendicular to $\vvorb$. Velocity kicks parallel to $\vvorb$ (black curves) -- which inject orbital energy -- also induce secular motion parallel to $\vvorb$. For any plausible Galactic potential, this secular motion is opposite the direction of the kick $\Delta\vv$, reflecting the notion that higher energy orbits have longer periods.\footnote{This phenomenon has been humorously dubbed the `donkey effect' owing to \citet{lynden1972generating}.}

Na\"ively computing the position perturbation, $\delta \vx$, as the difference in position between the perturbed and unperturbed orbits leads to the conclusion that the amplitude of oscillations in the component $\delta x_\parallel$ along $\vvorb$ can grow over time (dotted curves in the left-hand panel of Fig.~\ref{fig:dx}). However, this growth actually represents the uniform expansion and contraction of (non-self-bound) coorbiting material in proportion with its orbital velocity. The separation between orbiting objects oscillates in proportion to $\vorb$.\footnote{Two objects on the same orbit can be viewed as being separated in time by some static $\Delta t$. Their spatial separation is therefore $\Delta x = (\Delta x/\Delta t)\Delta t \to v\Delta t$ in the limit that $\Delta t$ is small. Nearby points on a stellar stream can be approximated as inhabiting the same orbit even if the entire stream cannot, so by this argument their spatial separation is proportional to their orbital velocity $\vorb$.} To factor out this effect, we define the position perturbation along $\vvorb$ to be
\begin{equation}\label{pertframe}
    \delta x_\parallel = (\vx-\vx_0)\cdot\uvvorb \frac{\langle\vorb\rangle}{\vorb},
\end{equation}
where $\vx$ is the perturbed position, $\vx_0$ is the unperturbed position, and $\langle\vorb\rangle$ is the time-averaged orbital velocity. The perturbation $\delta x_\parallel$ defined in this way is plotted in the left-hand panel of Fig.~\ref{fig:dx} (solid curves), where it is evident that the $1/\vorb$ rescaling has eliminated the growth in the oscillation amplitude. After many orbital periods, the oscillations in $\delta x_\parallel$ are much smaller than the secular component, so in our analysis we neglect the oscillatory motion and set
\begin{align}\label{dx}
    \delta x_\parallel &= \Lambda \Delta v_\parallel t,
\end{align}
where $\Delta v_\parallel$ is the component of the velocity kick parallel to $\vvorb$ and $t$ is the time elapsed since the kick. For the same reason, we neglect perturbation components $\delta x_\perp$ perpendicular to $\vvorb$. The parameter $\Lambda<0$, which is the mean slope of the solid black curve in the left-hand panel of Fig.~\ref{fig:dx}, is a function of the time of the kick; it varies over an orbital period because the efficiency of a kick depends on the instantaneous orbital velocity.\footnote{For instance, the energy injection is $\Delta E=\vvorb\cdot\Delta v$, so velocity kicks are most efficient at pericentre when $\vorb$ is highest.} In our simplified analysis, we treat $\Lambda$ as a random variable.

One problem remains. Due to the transformation in Eq.~(\ref{pertframe}), the relationship between distance scales along the stream and distance scales associated with the substructure environment varies over the orbital period. Within the stream's co-scaling frame, a velocity injection accumulated over many orbital periods becomes
\begin{align}\label{Dv_rand}
     \Dv^\textrm{co-}(\vr) \equiv \int\diff \omega f_\mathrm{orb}(\omega) \Dv(\omega \vr),
\end{align}
where $\omega=\vorb/\langle\vorb\rangle$ if $\vr$ lies along $\vvorb$.
In general we will assume that the separation vector $\vr$ between different points on the stream lies along $\vvorb$,\footnote{That is, we assume different parts of the stream lie on the same orbit. This approximation can become inaccurate over scales comparable to the size of the stream \citep[][]{sanders2013stream}.} so $f_\mathrm{orb}$ is indeed the distribution of $\vorb/\langle\vorb\rangle$. To give some context to the number $\omega=\vorb/\langle\vorb\rangle$, at pericentre $\omega>1$ (the stream is stretched), while at apocentre $\omega<1$ (the stream is compressed).

\subsection{Impact on the stream power spectrum}\label{sec:orbit_impact}

We now show how the power spectra derived in Sections \ref{sec:inhomogeneity} and~\ref{sec:disp} are modified by the approximate treatment of orbital dynamics encapsulated in Eqs. (\ref{dx}) and~(\ref{Dv_rand}).

\subsubsection{Velocity kick efficiency}

It is evident from Eq.~(\ref{dx}) that $\Lambda$ scales the efficiency of velocity kicks. That is, over many orbital periods, the response of a star to a velocity kick is as if the kick were $\Lambda$ times stronger (or weaker). Accordingly, the velocity-injection power spectrum derived in Section~\ref{sec:inhomogeneity} should be scaled by $\langle \Lambda^2 \rangle$ to account for this effect. Note that as a consequence, the diffusion coefficient $D$ (Section~\ref{sec:D}) is also scaled by the same factor.

\subsubsection{Distance transformation}

To account for the distance transformation in Eq.~(\ref{pertframe}), we can use Eq.~(\ref{Dv_rand}) to write the velocity-injection correlation function
\begin{align}
     \langle\Delta v_i^\textrm{co-}(0)\Delta v_i^\textrm{co-}(\vr)\rangle &=
     \int\diff \omega f_\mathrm{orb}(\omega) 
     \langle\Delta v_i(0)\Delta v_i(\omega \vr)\rangle
\end{align}
along any direction $i$. Consequently, the velocity-injection power spectrum becomes
\begin{align}\label{Ptrans}
    P_{\Delta v}^\textrm{co-}(k) = \int\diff \omega \frac{f_\mathrm{orb}(\omega)}{\omega} P_{\Delta v}(k/\omega)|_{v=\omega\langle \vorb\rangle}
\end{align}
in the frame that co-expands or contracts with the stellar stream. Recall that the velocity-injection power spectrum is sensitive to the stream velocity $v$, so in Eq.~(\ref{Ptrans}) we set $v=\omega\langle\vorb\rangle$. To transform the resulting stream power spectrum back to physical space, we may write
\begin{align}\label{Ptrans_inv}
    \Ps(k) = \Ps^\textrm{co-}(\omega_0 k),
\end{align}
where $\omega_0$ is the value of $\omega$ at the observation time.

We have assumed for simplicity that $\omega$ is not correlated with $\Lambda$, but one can straightforwardly account for such correlations by including the factor $\langle\Lambda^2|\omega\rangle$ (the expectation of $\Lambda^2$ given a particular value of $\omega$) inside the integrand in Eq.~(\ref{Ptrans}). Within the context of Eq.~(\ref{Ptrans}) it is also possible to allow the substructure power spectrum or velocity distribution to vary as a function of Galactocentric radius by varying these quantities as a function of $\omega$ within the expression for $P_{\Delta v}$; $\omega$ is set in turn by the Galactocentric radius (or vice versa). We do not pursue this possibility, but it represents another straightforward extension.

\section{Stream considerations}\label{sec:stream}

So far, we have considered the stellar stream as a collection of stars whose unperturbed motion is simply time-separated along a single orbit. However, in practice, even an unperturbed stream undergoes non-trivial global evolution. It forms from the continuous disruption of a star cluster or dwarf galaxy, and the velocity dispersion of the stripped material causes the stream to expand over time. We discuss in this section how these considerations impact our analytic treatment.

\subsection{Initial velocity dispersion}\label{sec:sigmaeff}

The analytic description of stream perturbations in Section~\ref{sec:disp} requires knowledge of the stream's initial velocity dispersion $\sigma_0$. In principle, as long as $Dt\ll\sigma_0^2$ it is possible to measure $\sigma_0$ directly from the velocities of stream stars. We warn, however, that $\sigma_0$ is \textit{not} the dispersion of instantaneous stellar velocities. Rather, it is the dispersion of net stellar displacements, per unit time, averaged over many orbital periods. Orbital dynamics make the connection between the instantaneous velocity dispersion and $\sigma_0$ non-trivial; a study of orbital velocity perturbations (similar to the study of position perturbations in Section~\ref{sec:orbit}) is necessary. We do not pursue such a study in this work.

If $\sigma_0$ cannot be measured directly, it can still be constrained using knowledge of the stream's global properties, such as its length and age. For instance, \citet{eyre/binney:2011} and \citet{bovy2014dynamical} connect a stream's unperturbed global evolution to the velocity dispersion $\sigma_\mathrm{p}$ of its progenitor cluster. Stripped stars are taken to escape the progenitor with velocities spread by $\sigma_\mathrm{p}$ about the progenitor's orbital velocity. Since stars are predominantly stripped near the cluster's pericentre, the velocities at which they escape are maximally efficient at setting their resulting displacements along the stellar stream. Thus, in the language of Section~\ref{sec:orbit} the resulting velocity dispersion of stream stars is
\begin{equation}
    \sigma_\mathrm{tot} \simeq (\max|\Lambda|)\sigma_\mathrm{p}
\end{equation}
with $\Lambda$ defined in Eq.~(\ref{dx}).

The total velocity dispersion $\sigma_\mathrm{tot}$ of stream stars is not, however, the same as the local velocity dispersion $\sigma_0$ that is relevant to the treatment in Section~\ref{sec:disp}. Instead, $\sigma_0<\sigma_\mathrm{tot}$ due to self-sorting wherein stars of similar velocities tend to arrive at similar positions. While $\sigma_0$ could vary over the length of the stream, for simplicity we write
\begin{equation}\label{sigmaeff}
    \sigma_0 = c\sigma_\mathrm{tot}
\end{equation}
with $c$ a constant. We estimate in Appendix~\ref{sec:sigmalocal}, using a idealized argument, that $c\simeq 0.4$ if the total velocity distribution is Maxwellian.

\subsection{Position-dependent age}\label{sec:age}

Because a stellar stream is continuously sourced, locations more distant from the progenitor have had more time to evolve in response to perturbations. The dynamical age of a point on the stream is proportional to its distance from the progenitor. This means that if we are studying the power spectrum of the whole stream, we are effectively averaging over ages $t$ of stream segments up to the total age $t_\mathrm{age}$ of the stream. At leading order, the power spectrum becomes
\begin{align}
    \Ps^\mathrm{avg}(k,t_\mathrm{age})=\frac{1}{t_\mathrm{age}}\int_0^{t_\mathrm{age}}\diff t \Ps(k,t).
\end{align}

\subsection{Initial density variations}\label{sec:di}

Even without perturbations, a stellar stream exhibits density variations arising from epicyclic oscillations of stream stars \citep[][]{kupper2010tidal,kupper2012more}. A variable rate of tidal stripping also contributes \citep{sanders/etal:2016}; most tidal stripping occurs near pericentre. In our analysis, this effect could be treated as an initial perturbation to the stream's distribution function within the framework of Section~\ref{sec:disp}. However, we expect that the velocity dispersion will rapidly suppress these perturbations, a view that aligns with the results of \citet{sanders/etal:2016} \citep[but for a counterpoint, see][]{ibata2020detection}. In any event, we neglect these epicyclic density variations in the present analysis, but we anticipate that the treatment in Section~\ref{sec:disp} can be straightforwardly extended to include them.

\section{Discussion}\label{sec:disc}

With the treatment of stellar streams complete, we now put it into practice and discuss some of its predictions and implications. Section~\ref{sec:procedure} discusses how to put the analytic treatment into practice, Section~\ref{sec:compare} compares analytic predictions to numerical simulations, and Section~\ref{sec:implications} discusses some of the analytic framework's implications.

\subsection{Putting the procedure together}\label{sec:procedure}

Application of the straightforward analytic formulae of Sections \ref{sec:inhomogeneity} and~\ref{sec:disp} is complicated by the array of practical refinements discussed in Sections \ref{sec:orbit} and~\ref{sec:stream}. Here we list the order in which we carry out each calculation.
\begin{enumerate}
    \item Integrate the stream's orbit to obtain the distributions of the perturbation coefficients $\Lambda$ and $\omega$ discussed in Section~\ref{sec:orbevo}.
    \item Approximate the stream's local initial velocity dispersion $\sigma_0$ as described in Section~\ref{sec:sigmaeff}.
    \item Compute the velocity-injection power spectrum $P_{\Delta v,\parallel}(k)$ as described in Section~\ref{sec:inhomogeneity}, transform it into the stream's co-scaling frame using Eq.~(\ref{Ptrans}), and scale it by $\langle\Lambda^2\rangle$.
    \item Evaluate the diffusion coefficient $D$ by integrating $P_{\Delta v,\parallel}(k)$ as described in Section~\ref{sec:D}.
    \item Compute the stream power spectrum $\Ps(k)$ as described in Section~\ref{sec:densitypower}.
    \item Average the result over stream ages as discussed in Section~\ref{sec:age}.
    \item Apply the inverse distance transformation, Eq.~(\ref{Ptrans_inv}).
\end{enumerate}

\subsection{Comparison to stream simulations}\label{sec:compare}

We now test the results of our analytic framework against numerical simulations. In particular, we compare stellar stream simulations executed using \textsc{galpy}'s stream-modelling functionality \citep{bovy2014dynamical,bovy2015galpy} in which stochastic substructure-induced perturbations are added by the \textsc{streampepperdf} extension \citep{bovy2017linear}. For simplicity, we consider the exact same scenario as \citet{bovy2017linear}, the details of which follow. The Galactic potential is modelled as a logarithmic potential with circular velocity 220~km\,s$^{-1}$ but is flattened by the factor 0.9 along one axis, making it axisymmetric. The stream is intended to resemble the GD-1 stream at the present time; its progenitor's orbit is described in \citet{bovy2014dynamical}\footnote{In particular, the stream progenitor's position and velocity today are taken to be $(x,y,z)=(12.4,1.5,7.1)$ kpc and $(v_x,v_y,v_z)=(107,-243,-105)$~km\,s$^{-1}$, where $z$ is the potential's symmetry axis.} and its progenitor's velocity dispersion is set to be $\sigma_p=(0.365\text{ km\,s}^{-1})(4.5\text{ Gyr}/t_\mathrm{age})$, where the stream age $t_\mathrm{age}$ remains free. We consider only the stream's leading arm, and these parameters give the arm a length today of about 12~kpc, of which its highest density portion occupies only 6~kpc. Substructure is taken to have the mass function $\diff n/\diff M\propto M^{-2}$ normalized such that the number density of subhaloes between $10^6$~$\mathrm{M}_\odot$ and $10^7$~$\mathrm{M}_\odot$ is $5.86\times 10^{-4}$ kpc$^{-3}$. Subhaloes are taken to have Hernquist density profiles (Eq.~\ref{hernquist}) with scale radius $R=1.05\text{ kpc} (M/10^8~\mathrm{M}_\odot)^{1/2}$.\footnote{We retain \citet{bovy2017linear}'s power-law form of the $R(M)$ relationship for simplicity, but the radius--mass (or equivalently, concentration--mass) relationship for field haloes is not a power law except at the largest mass scales \citep[e.g.][]{ludlow2016mass}, and the same is expected to hold for subhaloes. Note that $R\propto M^{1/2}$ corresponds to $c\propto M^{-1/6}$, where $c\propto M^{1/3}/R$ is the halo concentration.} Finally, the substructure velocity distribution is taken to be Maxwellian with scale velocity $u_0=120$~km\,s$^{-1}$.

To apply the analytic framework to this problem, we must first obtain the orbital perturbation parameters discussed in Section~\ref{sec:orbit}. By integrating the stream progenitor's orbit, we find that $\max|\Lambda|\simeq 1.27$ and $\langle\Lambda^2\rangle\simeq 0.95$.\footnote{Recall $\Lambda$ is the orbit-averaged value of $\delta x/(\Delta v t)$ for any given velocity kick. As we noted in Section~\ref{sec:orbevo}, $\Lambda<0$ in general for any realistic halo potential, and for a circular orbit in a logarithmic potential, $\Lambda=-1$ identically.} $\omega\equiv \vorb/\langle\vorb\rangle$ ranges from $0.70$ to $1.38$ with a present-day value of $\omega_0=1.33$ and an rms spread of about 0.22 from the mean of (by definition) 1, although when evaluating the velocity-injection power spectrum, we will use the full distribution of $\omega$ (Monte-Carlo sampled) and not its summary statistics. We also need the stream's time-averaged orbital velocity $\langle\vorb\rangle\simeq 215$~km\,s$^{-1}$. Finally, in estimating the diffusion coefficient $D$, we take the length of the stream to be $L=12$~kpc.

\begin{figure*}
	\centering
	\includegraphics[width=\linewidth]{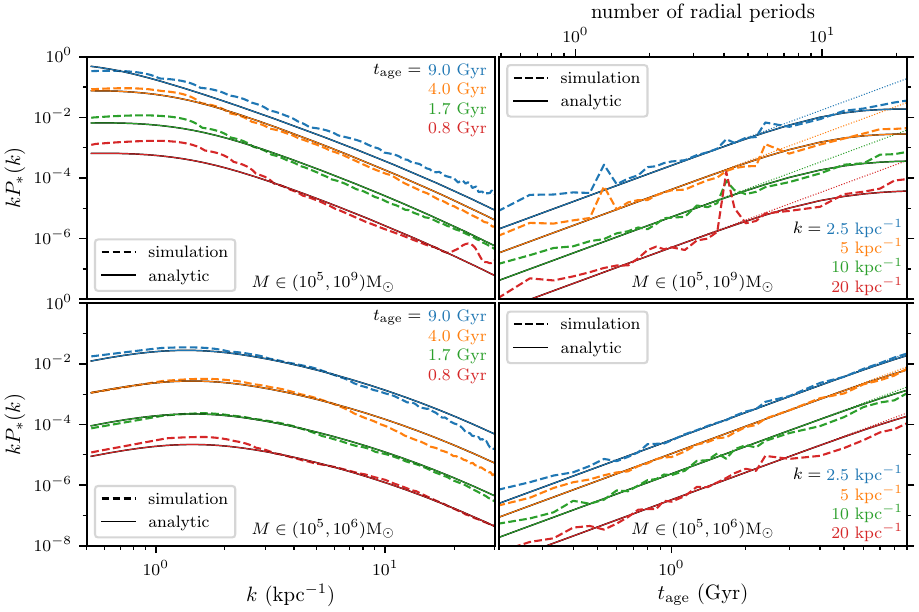}
	\caption{Comparison between simulated stream density power (dashed curves) and analytic predictions (solid curves). In the left-hand panels we plot the power spectrum for streams of several different ages $t_\mathrm{age}$, while on the right we show the density power at a few set wavenumbers $k$ as a function of $t_\mathrm{age}$ (or on the top axis, the number of radial orbit periods $t_\mathrm{age}/T_r$). Upper panels consider perturbations due to substructure in the mass range $(10^5,10^9)\mathrm{M}_\odot$, while the lower panels limit subhaloes to the mass range $(10^5,10^6)\mathrm{M}_\odot$. Within the left-hand panels, the power spectra are comparatively featureless except near $k=(\sigma_0 t_\mathrm{age})^{-1} \simeq 0.9$~kpc$^{-1}$, owing to our power-law assumptions about the subhalo spectrum. Note that this characteristic wavenumber is the same for all $t_\mathrm{age}$ values since $\sigma_0\propto t_\mathrm{age}^{-1}$ (see the text). Within the right-hand panels, power increases with age except when $t_\mathrm{age}\gtrsim \sigma_0^2/D\simeq 3.7$~Gyr in the upper panel ($D$ is negligibly small in the lower panel). We also show (thin dotted curves) the analytic predictions if the diffusion coefficient $D$ (and hence the induced velocity dispersion) is neglected (see Section~\ref{sec:D}). Note that the right-hand panels do \textit{not} depict time evolution because we vary the stream's initial velocity dispersion with its age; each point represents a different stream. The simulation curves on the right are noisy for the same reason: independent stream simulations randomly encounter more or less disruptive substructure. Broadly, despite the approximations we made in Sections \ref{sec:orbit} and~\ref{sec:stream}, our analytic predictions agree well with the simulation results except in a few regimes that we discuss in the text.}
	\label{fig:compare}
\end{figure*}

We carry out the simulation described above for two different substructure scenarios. In the first, we include subhaloes within the mass range $(10^5,10^9)\mathrm{M}_\odot$, while in the second, we consider only the subset of haloes that lie in the mass range $(10^5,10^6)\mathrm{M}_\odot$. We also explore a range of stream ages $t_\mathrm{age}$. For each simulation, we compute the power spectrum of the fractional contrast between the substructure-perturbed and unperturbed linear stream density. As in \citet{bovy2017linear}, we window the density contrast using a Hann window function to suppress edge effects in the power spectrum. We repeat each simulation at least 240 times with substructure encounters randomized and average the resulting power spectra. Figure~\ref{fig:compare} shows a sample of these simulation-averaged power spectra (dashed lines; left-hand panels) together with the corresponding analytic predictions (solid lines). Given the approximations we made in Sections \ref{sec:orbit} and~\ref{sec:stream}, the analytic predictions match the simulation results remarkably well. Recall that there are no tunable parameters in our analytic prediction, apart from the choice of cut-off in the computation of $D$ (which we turn to below). Generally, discrepancies arise at short $t_\mathrm{age}$, long $t_\mathrm{age}$, and large scales. We discuss these discrepancies next. There are also discrepancies at small scales for which we do not have an explanation, but we suspect that the simulation results may be unreliable in this regime due to finite spatial resolution.\footnote{In particular, we find that the simulation power spectra at $k\gtrsim 10$ kpc$^{-1}$ are not converged with respect to changes in the stream's angular resolution (in action-angle variables). In any event, these scales most likely lie below the Poisson noise floor, which would make them observationally irrelevant (see Fig.~\ref{fig:vary}).}

At large scales, several assumptions made in the analytic derivations break down. First, perturbations in this regime can arise from a small number of encounters with the largest haloes. For instance, 1~kpc corresponds to the scale radius of a $10^8$~$\mathrm{M}_\odot$ halo. If $t_\mathrm{age}=9$~Gyr, then about 63 halos are expected to pass within $5R$ of the stream, where $R$ is the halo's scale radius, but only one of these is expected to have a mass larger than $10^8$~$\mathrm{M}_\odot$. We will see in Section~\ref{sec:accuracy} that these close encounters contribute dominantly to the stream perturbations. In this case we have departed from the diffusion regime on which our analytic treatment is predicated. We remark that the match is much tighter at large scales between simulated and analytic power spectra in the $(10^5,10^6)\mathrm{M}_\odot$ substructure scenario than in the $(10^5,10^9)\mathrm{M}_\odot$ scenario (compare upper and lower panels in Fig.~\ref{fig:compare}), which suggests that this is the primary source of the large-scale discrepancy. However, we note that another assumption that breaks down is the approximation of the stream as a straight line. There is also a third effect that could harm the accuracy of the analytic predictions at large scales: the initial density variations discussed in Section~\ref{sec:di} that we neglected. However, this effect would not cause a discrepancy with the simulations because the simulations also neglect it.

We next discuss how the match between simulated and analytic power spectra depends on the stream age $t_\mathrm{age}$. To make this discussion clearer, we also plot in Fig.~\ref{fig:compare} (right-hand panels) the power at fixed wavenumber $k$ as a function of the stream age $t_\mathrm{age}$.\footnote{Recall that the stream's velocity dispersion $\sigma_0$ depends on $t_\mathrm{age}$ (in order to fix the stream's length today), so plotting the power spectrum as a function of $t_\mathrm{age}$ does \textit{not} produce a time-evolution plot. Each different $t_\mathrm{age}$ represents a simulation of a different stream.} The radial orbit period of the stream progenitor is about $T_r=0.41$~Gyr, and the treatment of orbit perturbations in Section~\ref{sec:orbit} assumes that $t_\mathrm{age}\gg T_r$. Surprisingly, however, the analytic predictions already match simulation results for $t_\mathrm{age}\gtrsim 1.5T_r$; a large number of orbits is not needed.

When $t_\mathrm{age}$ is long, on the other hand, the analytic predictions can fail because of the assumption that the diffusion coefficient $D$ is spatially uniform (see Section~\ref{sec:D}). If $Dt\gtrsim \sigma_0^2$, this assumption means that the velocity dispersion will be overestimated at small scales, and thus the power spectrum will be underestimated. In the $(10^5,10^9)\mathrm{M}_\odot$ substructure scenario, Fig.~\ref{fig:compare} shows that the analytic prediction accordingly begins to underestimate the power spectrum when $t_\mathrm{age}\gtrsim 7$~Gyr. We note that $Dt_\mathrm{age}=4.1\sigma_0^2$ for $t_\mathrm{age}=6$~Gyr and $Dt_\mathrm{age}=6.5\sigma_0^2$ for $t_\mathrm{age}=7$~Gyr, so $Dt/\sigma_0^2$ can become moderately greater than 1 before analytic predictions begin to significantly underestimate the power spectrum. In contrast, the $(10^5,10^6)\mathrm{M}_\odot$ substructure scenario yields $Dt_\mathrm{age}<\sigma_0^2$ for all ages we consider, so the analytic predictions here do not begin to diverge from simulation results at large $t_\mathrm{age}$. For comparison, Fig.~\ref{fig:compare} also shows (right-hand panel; thin dotted curves) the analytic predictions if we set $D=0$. As we suggested in Section~\ref{sec:D}, the predictions with $D=D(2\pi/L)$ and $D=0$ bracket the simulation result at large $t_\mathrm{age}$.

\subsection{Consequences}\label{sec:implications}

We now explore some of the implications of the analytic formulation of stellar stream perturbations.

\subsubsection{Time evolution of perturbations; impact of the velocity dispersion}\label{sec:timeevo}

\begin{figure}
	\centering
	\includegraphics[width=0.6867\linewidth]{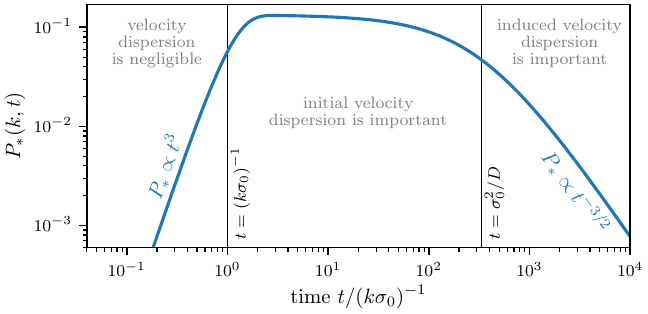}
	\caption{Illustration of the general time evolution of the stream density power spectrum $\Ps(k,t)$ at fixed wavenumber $k$. For this illustration we set $D=0.003k\sigma_0^3$. This evolution exhibits three main behaviours. When $k\sigma_0 t\ll 1$, the velocity dispersion is negligible, and the power spectrum grows as $\Ps\propto t^3$ (i.e. $\dot\delta_* \propto k\Delta v$ with $\Delta v\propto t^{1/2}$). When $k\sigma_0 t\gtrsim 2$, the power spectrum enters a steady state where the injection of new power is cancelled by suppression due to the constant velocity dispersion $\sigma_0$. Finally, when $t$ starts to approach $\sigma_0^2/D$ -- which implies the injected velocity dispersion $Dt$ becomes significant -- growth of the velocity dispersion $\sigma_0^2+Dt$ causes the power spectrum to suffer net suppression.}
	\label{fig:timeevo}
\end{figure}

Equation~(\ref{powerd}) describes the general time evolution of stream density perturbations, and this time evolution is illustrated in Fig.~\ref{fig:timeevo}. As we discussed in Section~\ref{sec:densitypower}, $\Ps(k,t)$ evidently goes through three distinct phases in its time evolution:
\begin{enumerate}
    \item Initially, when $t\ll(k \sigma_0)^{-1}$, the impact of the velocity dispersion is negligible and $\Ps(k,t)\propto t^3$ (Eq.~\ref{earlypower}).
    \item When $t\gtrsim 2(k\sigma_0)^{-1}$, the power spectrum enters a steady state wherein the (initial) velocity dispersion suppresses power at the same rate that new power is injected (Eq.~\ref{staticpower}).
    \item Eventually, velocities injected by substructure encounters start to significantly raise the stream's velocity dispersion. This effect causes velocity dispersion-induced suppression to ultimately outpace the injection of new power, so the density power decays. The characteristic time-scale for this effect is $\sigma_0^2/D$, but as Fig.~\ref{fig:timeevo} shows, the transition from the steady state into the decaying regime is very gradual. When $t\gg \sigma_0^2/D$, $\Ps(k,t)\propto t^{-3/2}$ (Eq.~\ref{decaypower}).
\end{enumerate}
It is possible to skip the steady-state phase if $D$ is sufficiently large, but in practice this is unlikely except at the largest scales. Because of the relationship $L\propto \sigma_0 t$ between the stream's length $L$ and its initial velocity dispersion, the power spectrum at any relevant wavenumber $k>2\pi/L$ enters the steady-state regime rapidly.

\begin{figure*}
	\centering
	\includegraphics[width=\linewidth]{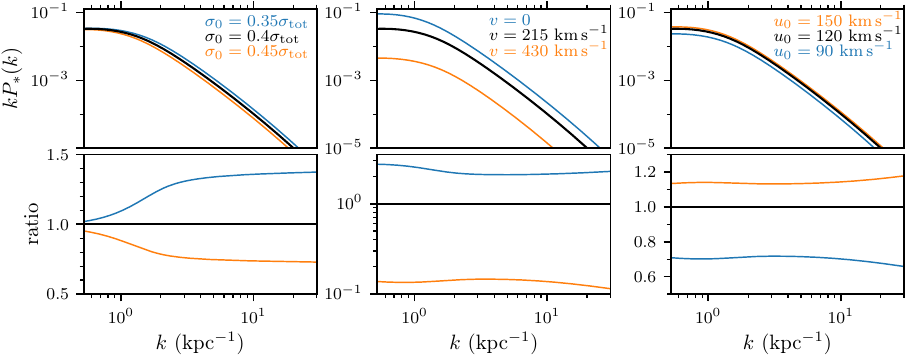}
	\caption{Varying the stream velocity dispersion $\sigma_0$, stream orbital velocity $v$, and substructure Maxwellian scale velocity $u_0$. We consider the same scenario as in Section~\ref{sec:compare}, but we set the stream's age to be $t_\mathrm{age}=3$~Gyr. Upper panels show the power spectrum $\Ps$ of stream density perturbations, while lower panels show the ratio of $\Ps$ to the power spectrum in the reference scenario.
	\textit{Left-hand panels}: the impact of the stream velocity dispersion $\sigma_0$. We suggested the approximation $\sigma_0=0.4\sigma_\mathrm{tot}$ in Section~\ref{sec:sigmaeff} by reference to Fig.~\ref{fig:sigma}, where $\sigma_\mathrm{tot}$ is the stream's total (global) velocity dispersion. Here we show the impact of that choice. Because $\Ps\propto \sigma_0^{-3}$ in the steady-state regime (Eq.~\ref{staticpower}), the approximately $13$ per cent changes to $\sigma_0$ that we consider lead to 40 per cent changes in $\Ps(k)$ at high $k$.
	\textit{Centre panels}: the impact of the stream's orbital velocity $v$. Fiducially, we take $v=215$~km\,s$^{-1}$. Recall that $u_0=120$~km\,s$^{-1}$, so we vary $v$ between $0$ and $3.6 u_0$. The resulting change in $\Ps$ is substantial.
	\textit{Right-hand panels}: the impact of the scale velocity $u_0$ of the substructure's Maxwellian velocity distribution. Despite the $1/u_0$ factor in the expression for the velocity-injection power spectrum $P_{\Delta v}$ (e.g. Eq.~\ref{Pvpara_MB}), the exponential factor wins, causing larger $u_0$ to yield larger $\Ps$.
	}
	\label{fig:vary_v}
\end{figure*}

Within the steady-state regime, Eq.~(\ref{staticpower}) implies that $\Ps\propto \sigma_0^{-3}$. In Section~\ref{sec:sigmaeff}, we suggested the one-significant-figure approximation $\sigma_0\simeq 0.4\sigma_\mathrm{tot}$, and the strong sensitivity of $\Ps$ to $\sigma_0$ suggests that this approximation may be cause for concern. Figure~\ref{fig:vary_v} (left-hand panels) shows the impact of choosing $\sigma_0=0.35\sigma_\mathrm{tot}$ or $\sigma_0=0.45\sigma_\mathrm{tot}$ instead. These approximately $13$ per cent changes to $\sigma_0$ lead to 40 per cent changes in $\Ps(k)$ at high $k$. In practice, however, the argument in Appendix~\ref{sec:sigmalocal} only motivates a value of $\sigma_0$ roughly ranging from $0.380\sigma_\mathrm{tot}$ to $0.415\sigma_\mathrm{tot}$. We also remark that if the (initial) local velocity dispersion $\sigma_0$ can be measured directly from the velocities of stream stars, as we discuss in Section~\ref{sec:sigmaeff}, then the value of $\sigma_0/\sigma_\mathrm{tot}$ is irrelevant.

\subsubsection{Impact of the substructure power spectrum}

Suppose the substructure power spectrum is a power law, $\mathcal{P}(q)\propto q^{c+3}$ or $P(q)\propto q^{c}$. This power spectrum could arise, for example, from a halo population with mass function $\diff n/\diff M\propto M^{-\alpha}$ and radius--mass relation $R\propto M^\beta$; then $c=(\alpha-3)/\beta$ (see Eq.~\ref{Pksub}). More generally, a power law is usually a good approximation over a limited range of wavenumbers. If in particular
\begin{equation}
    \mathcal{P}(q) = \mathcal{P}_0\, (q/q_0)^{c+3},
\end{equation}
then Eq.~(\ref{Pvpara_MB}) implies that
\begin{equation}
    P_{\Delta v,\parallel}(k) = (2\pi)^{7/2}\mathcal{P}_0
    \frac{G^2 \rhob^2 t}{q_0^4 u_0}\,
    \gamma\!\left(\frac{3-c}{2},\frac{v^2}{2 u_0^2}\right)
    \left(\frac{v}{\sqrt{2}u_0}\right)^{c-3}
    \left(\frac{k}{q_0}\right)^{c-1},
\end{equation}
where $\gamma(s,x)\equiv\int_0^x t^{s-1}\e^{-t}\diff t$ is the lower incomplete gamma function. This equation, together with Eqs.~(\ref{earlypower}--\ref{decaypower}), implies the scalings
\begin{align}\label{scaling1}
    P_{\Delta v,\parallel}(k)&\propto k^{c-1} \propto k^{-4}\mathcal{P}(k)
    \\
    \Ps(k)&\propto \left\{\begin{array}{cl}
    k^{c+1} \propto k^{-2}\mathcal{P}(k) & \text{in the growing regime,} \\[3pt]
    k^{c-2} \propto k^{-5}\mathcal{P}(k) & \text{in the steady-state and decaying regimes.}
    \end{array}\right.
    \label{scaling3}
\end{align}
For instance, the substructure scenario considered in Section~\ref{sec:compare} has $\mathcal{P}(q)\propto q$ (i.e. $c=-2$). The upper left panel of Fig.~\ref{fig:compare} appropriately depicts $k\Ps(k)\propto k^0$ at large scales $k\lesssim (\sigma_0 t_\mathrm{age})^{-1}\simeq 0.9$~kpc$^{-1}$, which are in the growing phase, and $k\Ps(k)\propto k^{-3}$ at small scales $k\gtrsim (\sigma_0 t_\mathrm{age})^{-1}$, which are in the steady-state or decaying phase.\footnote{In contrast to the upper left panel, the lower left panel of Fig.~\ref{fig:compare} depicts $k\Ps(k)$ growing with $k$ at large scales because the smaller $10^6$~$\mathrm{M}_\odot$ maximum halo mass in this scenario causes the substructure power spectrum to gradually transition into $\mathcal{P}(q)\propto q^3$ (Poisson noise), for which $k\Ps(k)\propto k^2$ in the growing phase. The same effect is visible in the right-hand panel of Fig.~\ref{fig:vary} for the $10^9$~$\mathrm{M}_\odot$ maximum halo mass scenario, but only at scales larger than the stream's length.} We also remark that even if the substructure power spectrum is not a power law, the steep factor $q^{-6}$ by which the integral in Eq.~(\ref{Pvpara}) suppresses the contribution from substructure modes $q>k$ (where $k$ is the wavenumber on the stream) implies that Eqs. (\ref{scaling1}) and~(\ref{scaling3}) remain approximately valid.

\subsubsection{Impact of the stream velocity}

We found in Section~\ref{sec:inhomogeneity} that the stream velocity $v$ limits the range of substructure scales $q$ that contribute to stream perturbations when $v$ exceeds the substructure velocity $\tilde u$ (in the Galactic frame); see Eq.~(\ref{Pvpara}). If the substructure velocity distribution is Maxwellian with scale velocity $u_0$, this effect resulted in an exponential factor $\exp[-(k^2/q^2)v^2/u_0^2]$ inside the integrand in Eq.~(\ref{Pvpara_MB}). Consequently, we expect that large stream velocities $v>u_0$ can significantly suppress perturbations to the stream. We explore this effect in the centre panels of Fig.~\ref{fig:vary_v}, where we vary the stream velocity between $v=0$ and $v\simeq 3.6 u_0$. Compared to the fiducial $v\simeq 1.8u_0$ used in Section~\ref{sec:compare}, setting $v=0$ more than doubles the stream power spectrum $\Ps$. Meanwhile, doubling the stream velocity to $v\simeq 3.6 u_0$ reduces $\Ps$ by nearly a factor of 10.

\subsubsection{Impact of substructure velocities}

The velocity-injection power spectrum contains the factor $1/u_0$ (see Eq.~\ref{Pvpara_MB}), where $u_0$ is again the scale velocity of the substructure's Maxwellian velocity distribution. This suggests that lower substructure velocities lead to greater stream power $\Ps$. However, altering $u_0$ also impacts velocity injections through the exponential factor $\exp[-(k^2/q^2)v^2/u_0^2]$, the effect of which trends in the opposite direction: reducing $u_0$ suppresses this factor. Figure~\ref{fig:vary_v} (right-hand panels) shows that -- at least for $v\gtrsim 1.5 u_0$ -- the latter effect wins, and increasing $u_0$ leads to increased $\Ps$. Interestingly, while one might na\"ively suspect that larger substructure velocities boost $\Ps$ because they lead to more encounters, mathematically that is not what occurs. Rather, in terms of the discussion in Section~\ref{sec:inhomogeneity}, the reduced substructure-encounter time-scale associated with larger $u_0$ allows substructure power over a broader range of scales to contribute to the inhomogeneous heating of the stream.

\subsubsection{Impact of subhalo concentrations}

\begin{figure*}
	\centering
	\includegraphics[width=\linewidth]{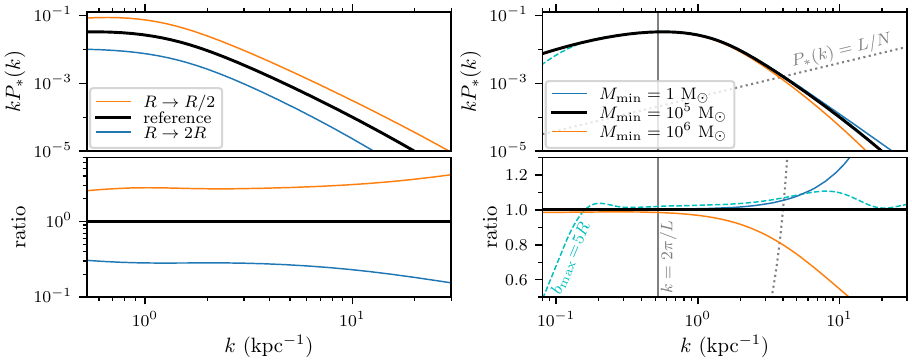}
	\caption{The impact of subhalo concentrations (left-hand panels) and the minimum subhalo mass (right-hand panels) on the power spectrum $\Ps$ of stream density perturbations. We consider the same scenario as in Section~\ref{sec:compare}, but we set the stream's age to be $t_\mathrm{age}=3$~Gyr. The upper panels show the stream power spectrum $\Ps$ in each scenario, while the lower panels show the ratio between $\Ps$ and the power spectrum in the reference scenario.
	\textit{Left-hand panels}: impact of the subhalo scale radii $R$. Raising (lowering) all $R$ by a factor of 2 causes $\Ps$ to decrease (increase) by a factor of about 3.
	\textit{Right-hand panels}: impact of the minimum subhalo mass $M_\mathrm{min}$. If we reduce $M_\mathrm{min}$ from $10^5$~$\mathrm{M}_\odot$ to $1$~$\mathrm{M}_\odot$ (blue curve), the resulting boost to $\Ps$ lies largely below the Poisson noise floor (dotted line), which we compute assuming the stream has length $L=12$ kpc and contains $N=3\times 10^4$ stars \citep[similar to the initial star count for GD-1 estimated in][]{de2020closer}. Consequently, subhaloes below $10^5$~$\mathrm{M}_\odot$ are likely undetectable using GD-1 density variations. On the other hand, raising $M_\mathrm{min}$ from $10^5$~$\mathrm{M}_\odot$ to $10^6$~$\mathrm{M}_\odot$ (orange curve) results in a suppression to $\Ps$ that is small, above the noise floor, but not necessarily unreachable.
	We also show (thin dashed cyan curve) the impact of the restriction made in the simulations of Section~\ref{sec:compare} and \citet{bovy2017linear} that only subhaloes that pass within $5R$ of the stream, where $R$ is the subhalo's scale radius, can perturb the stream. This restriction does not significantly harm the accuracy of the simulation's results because it only has a major impact on $\Ps$ at wavenumbers $k<2\pi/L$, i.e. scales larger than the length of the stream (but see also footnote~\ref{foot:bmax}).}
	\label{fig:vary}
\end{figure*}

Suppose we rescale all subhalo scale radii by some factor $a$, so $R\to aR$. Note that at fixed mass $M$, this means a subhalo's internal density scales by $a^{-3}$; if $a<1$ the haloes become more centrally concentrated, and vice versa if $a>1$. By Eq.~(\ref{Pksub}) the (dimensionful) substructure power spectrum simply shifts to $P(q)\to P(aq)$. Since $P(q)\propto q^{-2}$ (that is, $\mathcal{P}(q)\propto q$) for the substructure scenario under consideration, this results in the scaling $P(q)\to a^{-2}P(q)$ under the approximation that $P(q)$ is exactly a power law. Velocity injections and hence stream density perturbations then scale by the same factor $a^{-2}$, at least when induced velocity dispersions are negligible, i.e. $Dt \ll \sigma_0^2$. In particular, more concentrated haloes ($a<1$) result in larger stream density perturbations. This argument is approximate, and we show in Fig.~\ref{fig:vary} more precisely how the stream power spectrum $\Ps$ responds if subhalo radii $R$ are decreased (increased) by a factor of 2. Evidently, $\Ps$ grows (shrinks) by a factor of about $2^{1.6}\simeq 3$ in response.

\subsubsection{Contribution from low-mass subhaloes}

Within the simulations in Section~\ref{sec:compare} and \citet{bovy2017linear}, the minimum subhalo mass was set to $10^5$~$\mathrm{M}_\odot$. Inclusion of smaller halo masses would raise the cost of the simulation considerably; if $\diff n/\diff M\propto M^{-2}$, then each smaller decade in halo mass contains ten times more haloes than the last. However, using our analytic framework we can straightforwardly consider smaller halo masses. In fact, the approach becomes progressively more accurate when considering less massive and more numerous subhaloes. In the right-hand panel of Fig.~\ref{fig:vary}, we show the impact of reducing the minimum halo mass to $1$~$\mathrm{M}_\odot$. Evidently, the influence of subhaloes smaller than $10^5$~$\mathrm{M}_\odot$ on the stream power spectrum $\Ps(k)$ lies almost wholly below the Poisson noise floor for a stream comparable to GD-1. We also consider the impact of raising the minimum mass to $10^6$~$\mathrm{M}_\odot$. This change has a 10 to 20 per cent impact above the noise floor, suggesting that subhaloes below $10^6$~$\mathrm{M}_\odot$ could be detectable using this stream.

\subsubsection{Accuracy of finite simulations}\label{sec:accuracy}

In order to conserve computational expense, the simulations in Section~\ref{sec:compare}, and those in \citet{bovy2017linear}, limit the distances at which substructure-stream encounters are sampled. In particular, encounters are only considered whose impact parameter $b$ (with respect to the nearest point on the stream) satisfies $b<\bmax=5R$, where $R$ is the subhalo's scale radius. That is, the subhalo must pass within a distance $5R$ of the stream. We now test whether this restriction significantly alters the resulting stream density perturbations. For this purpose, we show in Appendix~\ref{sec:bmax} that if each star is only perturbed by substructure encounters with $b<\bmax$, then for the purpose of velocity injections, the substructure power spectrum is effectively scaled by $[1-J_0(\bmax k)]^2$, where $J_0$ is a Bessel function of the first kind. Note that this is a more severe restriction than that imposed by the simulations: we not only limit our subhaloes to those that pass within $\bmax$ of the stream, but additionally each star in the stream is only perturbed by subhaloes passing within $\bmax$ of that star. To make $\bmax$ a function of the subhalo's scale radius $R$, we insert the scaling factor inside the mass integral in Eq.~(\ref{Pksub}). The stream density power spectrum that results from this scaled substructure power spectrum $\Ps$ is shown (thin dashed cyan curve) in the right panel of Fig.~\ref{fig:vary}. The only significant change to $\Ps$ as a result of this restriction lies at wavenumbers $k<2\pi/L$, that is, scales larger than the length of the stream.\footnote{\label{foot:bmax} More generally, the restriction $b<5R$ has a major impact only on scales larger than the scale radius $R=3.3$~kpc associated with the maximum halo mass. It also causes a minor ringing effect -- partially visible as a bump at large $k$ in Fig.~\ref{fig:vary} -- close to the scale radius associated with the minimum halo mass.} Thus, the restriction that $b<\bmax=5R$ does not significantly harm the accuracy of the simulations.

\subsubsection{Impact of past environment}\label{sec:t}

We assumed in Sections \ref{sec:formalism} and~\ref{sec:inhomogeneity} that the perturbing environment's properties, such as its power spectrum and velocity dispersion, are fixed in time. However, a stream may have orbited the Galactic halo for the majority of the Galaxy's lifetime. To what extent do we need to worry that the substructure environment's properties in the past may differ from those today? The answer to this question lies within our derivation of the stream power spectrum $P_*(k,t)$ in Section~\ref{sec:densitypower}. In particular, $P_*(k,t)$ is written in Eq.~\eqref{powerd-time} as an integral over infinitesimal density power elements
\begin{equation}
    \diff P_*(k,t) = k^2 (t-t^\prime)^2
    \exp\!\left[-k^2 \left(\sigma_0^2+D \frac{t+2t^\prime}{3}\right)(t-t^\prime)^2\right]
    \step(t-t^\prime)
    \frac{\diff P_{\Delta v,\parallel}(k,t^\prime)}{\diff t'} \diff t'
\end{equation}
that each begin to grow due to a spectrum $\diff P_{\Delta v,\parallel}(k,t^\prime)$ of infinitesimal velocity kicks that occur at one time $t^\prime$. The density power element initially grows as $(t-t^\prime)^2$ before eventually suffering exponential suppression due to the stream's velocity dispersion. Due to this exponential suppression, $\diff P_*(k,t)$ has already fallen to about 3 per cent of its maximum value by the time $t-t^\prime\simeq 2.5/(k\sigma_0)$ in the case where the induced velocity dispersion $Dt$ is negligible. If $Dt$ is non-negligible, the suppression is even more extreme. Thus, we suggest that only the properties of the substructure environment within roughly the most recent time period $t_\mathrm{env}\equiv 2.5/(k\sigma_0)$ are relevant to stream perturbations at the scale $k$.

Let us explore briefly what this result means for the GD-1 stream models described in Section~\ref{sec:compare}. In these models $\sigma_0\simeq  (0.67~\mathrm{kpc})/t_\mathrm{age}$, so $t_\mathrm{env}/t_\mathrm{age} \simeq 3.7~\mathrm{kpc}^{-1}/k$. Thus for wavenumbers $k\lesssim 3.7~\mathrm{kpc}^{-1}$, the properties of the substructure environment over the entire past lifetime of the stream contribute non-trivially to its present-day density power spectrum. For $k\simeq 7~\mathrm{kpc}^{-1}$, only the environmental properties during the latter half of the stream's past lifetime play a significant role.

It is straightforward, in any event, to account explicitly for changes to the perturbing environment's properties. By considering a sequence of infinitesimal time periods during which the substructure properties are constant, we may allow the time derivative of the velocity-injection power spectrum $P_{\Delta v,\parallel}(k,t)$ to vary with time. For instance, from Eq.~(\ref{Pvpara_MB}) we obtain
\begin{align}\label{Pvpara_MB_t}
    \frac{\diff P_{\Delta v,\parallel}(k,t)}{\diff t} &=
    16\pi^4 G^2 k^2 \rhob^2\frac{\sqrt{2/\pi}}{u_0}
    \int_k^\infty\dqq\frac{\mathcal{P}(q)}{q^6}
    \exp\!\left(-\frac{1}{2}\frac{k^2}{q^2}\frac{v^2}{u_0^2}\right),
    &\text{(Maxwellian substructure)}
\end{align}
where $\rhob$, $\mathcal{P}(q)$, $u_0$, and $v$ may now be functions of time. Then one can substitute this expression into Eq.~(\ref{powerd-time}) to obtain the stream power spectrum.

\section{Conclusion}\label{sec:conclusion}

Stellar streams retain a strong memory of any past gravitational perturbations, a property that allows them to probe dark matter substructure that is otherwise invisible. In this article we developed a fully analytic description of the perturbation of a stellar stream by an inhomogeneous environment. This description rests on the assumption of many weak substructure encounters and has two main components: the dependence of integrated velocity injections on the environment, treated in Section~\ref{sec:inhomogeneity}, and the stellar stream's response, treated in Section~\ref{sec:disp}. In particular, we derive first how the power spectrum $\mathcal{P}(q)$ of the substructure environment sets the power spectrum $P_{\Delta v}(k)$ of velocity injections into the stream and next how the power spectrum $\Ps(k)$ of stream density perturbations responds to $P_{\Delta v}(k)$. Note that the substructure power spectrum $\mathcal{P}(q)$ can include contributions not only from dark matter but also from baryonic substructures such as giant molecular clouds and star clusters.

The analytic treatment laid out here describes an idealized one-dimensional system, which neglects the dynamics of the stream's orbit about the galaxy as well as the stream's global evolution. However, in Sections \ref{sec:orbit} and~\ref{sec:stream} we showed how the treatment can be applied, under certain approximations, to a realistic stellar stream. Here, the main approximations are that we average perturbations to stars' orbits over many orbital periods and we simplify the process by which a stream forms and grows. Approximations included, the full list of calculation steps is enumerated in Section~\ref{sec:procedure}. We compared in Section~\ref{sec:compare} the resulting analytic predictions to numerical simulations carried out using the methods (and code) of \citet{bovy2017linear}. Analytic predictions tend to fail in two main regimes: at the largest scales, which lie outside the diffusion regime, and when the stream has completed fewer than about 1.5 Galactic orbits. Otherwise, the agreement with simulations is remarkably tight.

Our analytic description reveals numerous insights about the behaviour of perturbed stellar streams, a key example of which is the time evolution of stream perturbations (Fig.~\ref{fig:timeevo}). These perturbations grow rapidly due to substructure interactions until the stream's velocity dispersion starts to significantly suppress them; at this point the perturbation spectrum enters a steady state wherein the suppression of power cancels the continuous injection of new power. The shape of the stream power spectrum $\Ps$ is in turn strongly influenced by the question of which scales are in the growing or steady-state regimes. The analytic description also clarifies the influence of myriad other factors on $\Ps(k)$. We discuss these results at length in Section~\ref{sec:implications}.

The speed and precision of our analytic formulation also makes it a useful forecasting tool. The GD-1 \citep{GD:2006} and Pal~5 \citep{odenkirchen2001detection} streams have been the primary targets of searches for substructure-induced perturbations. However, many more streams are known; see \citet{newberg2016tidal} for a review and 
\citet{Vickers_2015,
2017ApJ...847..119G,
Shipp_2018,
2018MNRAS.481.3442M,
2018ApJ...865...85I,
2019A&A...622L..13M,
2019ApJ...872..152I,
2019ApJ...886L...7M,
2019A&A...621L...2R,
necib2020evidence,
2021MNRAS.504.2727P,
2021MNRAS.507.1923J,
2021ApJ...920...51M,
2021ApJ...914..123I}
for more recent discoveries. In future work we will explore the prospects of other streams for probing dark matter substructure.

As a final remark, we note that stellar streams are not the only systems in which a dynamical response to dark matter substructure may be observed. Other examples include the disruption of binary star systems \citep{2010arXiv1005.5388P}, heating of globular clusters \citep{10.1093/mnras/stz2118}, heating of galactic discs \citep{10.1111/j.1365-2966.2004.07870.x,10.1093/mnras/stz534}, heating of the circumgalactic medium \citep{2020MNRAS.499.3255M}, perturbations to Solar System dynamics \citep{2013JCAP...03..001G}, and perturbations to the orbits of globular clusters or other objects within the Galactic halo \citep{10.1093/mnras/stz2648,10.1093/mnras/stab461}. The analytic velocity-injection formalism presented in Section~\ref{sec:formalism} is not specific to stellar streams; it is applicable to any system that is subjected to substructure encounters. A key advantage of our velocity-injection formalism over previous analytic treatments of substructure-induced dynamical perturbations is its direct formulation in terms of the substructure density field rather than a discrete-particle approximation.

\section*{Acknowledgements}

We thank Simon White, Sam Young, Elisa Ferreira, Ippei Obata, Minh Nguyen, Andrija Kostic, and Laura Herold for useful discussions. 
We also thank the anonymous referee whose detailed feedback helped to improve the clarity of this work significantly.
FS acknowledges support from the Starting Grant (ERC-2015-STG 678652) `GrInflaGal' from the European Research Council.

\section*{Data Availability}
 
No new data were generated or analysed in support of this
research.



\bibliographystyle{mnras}
\bibliography{main}



\appendix

\section{Applying the subfield decomposition}\label{sec:subfield_demo}

In Section~\ref{sec:general}, we presented the extension of the velocity-injection formalism in Section~\ref{sec:static} to a general substructure velocity distribution. This extension involves decomposing the density field $\delta(\vx)$ into subfields $\delta_j(\vx)$ that each have an associated velocity. We now illustrate how we use this construction to compute velocity-injection correlations. Starting from the velocity-injection expression, Eq.~(\ref{Dv_x}), we find that
\begin{align}
	\langle\Dv(0)\cdot\Dv(\vec r)\rangle
	&=
	\int\!\!\dnx \int\!\!\dnx^\prime \sum_{j,j^\prime=1}^N  \langle\delta_{j}(\vx)\delta_{j^\prime}(\vx^\prime)\rangle \vV(\vx|\vu_j,t)\cdot \vV(\vx^\prime-\vr|\vu_{j^\prime},t)
	&& \text{from the subfield decomposition (Eq.~\ref{delta_decomp}),}
	\\
	&=
	\int\!\!\dnx \int\!\!\dnx^\prime \sum_{j=1}^N  \langle\delta_{j}(\vx)\delta_{j}(\vx^\prime)\rangle \vV(\vx|\vu_j,t)\cdot \vV(\vx^\prime-\vr|\vu_{j},t)
	&& \text{using independence of subfields (Eq.~\ref{delta_independence}),}
	\\
	&=
	\int\!\!\dnq\int\!\!\dnqprime\sum_{j=1}^N \langle\delta_j(\vq)\delta_j^*(\vq^\prime)\rangle
	 \vV^*(\vq|\vu_j,t)\cdot \vV(\vq^\prime|\vu_j,t)\e^{-\I \vq^\prime\cdot\vr}
	&& \text{substituting inverse Fourier transforms,}
	\\\label{vcorr_step}
	&=
	\int\!\!\dnq\e^{-\I \vq \cdot\vr}\frac{P(q)}{N}\sum_{j=1}^N|\vV(\vq|\vu_j,t)|^2
	&& \begin{gathered}\text{by the power spectrum definition (Eq.~\ref{delta_independence})}\\\text{and the subfield power relation (Eq.~\ref{subpower}),}\end{gathered}
	\\
	&\to
	\int\!\!\dnq P(q)\e^{-\I \vq \cdot\vr}\int\diff^3\vu f(\vu) |\vV(\vq|\vu,t)|^2
	&& \text{in the continuum limit (Eq.~\ref{sum-to-v-integral}).}
\end{align}

\section{Velocity-injection power spectrum -- the substructure angular integral}\label{sec:angular}

Equation~(\ref{pow_dv}) expresses the one-dimensional power spectrum of velocity injections along two arbitrary unit vectors $\uvec a$ and $\uvec b$. In this appendix, we carry out the angular $\vq$ integrals in this expression. We first use the $\ddelta(\uvq\cdot\uvu)$ delta function to reduce the $\vq$ integral's dimensionality, rewriting Eq.~(\ref{pow_dv}) as
\begin{align}\label{Pab0}
	P_{\uvec a\uvec b}(k) 
	= 16\pi^3 G^2\rhob^2 t
    \int_0^\infty\dqq\frac{\mathcal{P}(q)}{q^3}
    \ints\dnu \frac{f(\vu)}{u}
    \int\diff\uvq_\perp
    \ddelta([\uvq_\perp\cdot\uvr]q+k)
    (\uvec a\cdot\uvq_\perp)(\uvec b\cdot\uvq_\perp).
\end{align}
Here, $\mathcal{P}(q)\equiv[q^3/(2\pi^2)]P(q)$ again and $\uvq_\perp$ is a unit vector integrated around a circle within the plane perpendicular to $\uvr$.

To carry out the $\uvq_\perp$ integral, we first establish a coordinate system. Let $\uvec p_1$ and $\uvec p_2$ be two unit vectors perpendicular to $\uvr$ and each other, so $\uvr$, $\uvec p_1$, and $\uvec p_2$ form an orthonormal basis. Also let $\eta\equiv \uvu\cdot\uvr$ and let $\psi$ be the angle from $\uvec p_1$ to the component of $\uvu$ perpendicular to $\uvr$. Then
\begin{equation}
    \uvu=\eta\uvr+\sqrt{1-\eta^2}\cos\psi\uvec p_1+\sqrt{1-\eta^2}\sin\psi\uvec p_2,
\end{equation}
i.e. we have constructed spherical polar coordinates for $\vu$ where $\eta$ is the cosine of the polar angle and $\psi$ is the azimuthal angle. If we next define $\phi$ to be the angle between $\uvq_\perp$ (which is perpendicular to $\uvu$) and the component of $\uvr$ perpendicular to $\uvu$, then
\begin{equation}\label{qperp}
    \uvq_\perp=
    \sqrt{1-\eta^2}\cos\phi\uvr
    +
    (-\eta\cos\psi\cos\phi+\sin\psi\sin\phi)\uvec p_1
    +
    (-\eta\sin\psi\cos\phi-\cos\psi\sin\phi)\uvec p_2
\end{equation}
and the integral over $\uvq_\perp$ becomes an integral over $\phi$. That is,
\begin{align}\label{Pab1}
	P_{\uvec a\uvec b}(k) 
	= 16\pi^3 G^2\rhob^2 t
    \int_0^\infty\dqq\frac{\mathcal{P}(q)}{q^3}
    \ints\dnu \frac{f(\vu)}{u}
    \int_0^{2\pi}\diff\phi
    \,\ddelta\!\left(\sqrt{1-\eta^2}\cos\phi\, q+k\right)
    (\uvec a\cdot\uvq_\perp)(\uvec b\cdot\uvq_\perp)
\end{align}
with $\uvq_\perp$ a function of $\phi$. Due to the argument of the delta function, the $\phi$ integral is only non-zero when $\sqrt{1-\eta^2}q>k$, in which case
\begin{equation}
    \ddelta\!\left(\sqrt{1-\eta^2}\cos\phi\, q+k\right)
    =
    \frac{\ddelta(\phi-\phi_+)}{\sqrt{1-\eta^2}\,|\sin\phi_+|q}
    +
    \frac{\ddelta(\phi-\phi_-)}{\sqrt{1-\eta^2}\,|\sin\phi_-|q},
\end{equation}
where $\phi_\pm$ are the two solutions to $\sqrt{1-\eta^2}\cos\phi\, q+k=0$ inside $0<\phi<2\pi$. In particular,
\begin{align}
    \cos\phi_\pm = \frac{-k}{q\sqrt{1-\eta^2}}
    \ \ \text{and}\ \ 
    \sin\phi_\pm = \pm\sqrt{1-\frac{k^2}{q^2(1-\eta^2)}}.
\end{align}
Consequently,
\begin{align}\label{Pab2}
	P_{\uvec a\uvec b}(k) 
	&= 16\pi^3 G^2\rhob^2 t
    \int_0^\infty\dqq\frac{\mathcal{P}(q)}{q^3}
    \ints\dnu \frac{f(\vu)}{u}
    \frac{\step(\sqrt{1-\eta^2}q-k)}{\sqrt{q^2(1-\eta^2)-k^2}}
    \left\{
    \left.(\uvec a\cdot\uvq_\perp)(\uvec b\cdot\uvq_\perp)\right|_{\phi=\phi_+}
    +
    \left.(\uvec a\cdot\uvq_\perp)(\uvec b\cdot\uvq_\perp)\right|_{\phi=\phi_-}
    \right\},
\end{align}
where $\step$ is the (Heaviside) unit step function.

Without loss of generality, we focus now on the cases where $\uvec a$ and $\uvec b$ take the basis values $\uvr$, $\uvec p_1$, or $\uvec p_2$. The general power spectrum is simply a bilinear combination of such basis power spectra. Equations (\ref{qperp}) and~(\ref{Pab2}) imply
\begin{align}\label{Pab3}
	P_{\uvr\uvr}(k) 
	&= 32\pi^3 G^2\rhob^2 k^2 t
    \int_0^\infty\dqq\frac{\mathcal{P}(q)}{q^6}
    \ints\dnu \frac{f(\vu)}{u}
    \frac{\step(1-\eta^2-k^2/q^2)}{\sqrt{1-\eta^2-k^2/q^2}}
\end{align}\begin{align}
	P_{\uvec p_1\uvec p_1}(k) 
	&= 32\pi^3 G^2\rhob^2 t
    \int_0^\infty\dqq\frac{\mathcal{P}(q)}{q^6}
    \ints\dnu \frac{f(\vu)}{u}
    \frac{\step(1-\eta^2-k^2/q^2)}{\sqrt{1-\eta^2-k^2/q^2}}
    \left[
    q^2\sin^2\psi - k^2\frac{\sin^2\psi-\eta^2\cos^2\psi}{1-\eta^2}
    \right]
\end{align}\begin{align}
	P_{\uvec p_2\uvec p_2}(k) 
	&= 32\pi^3 G^2\rhob^2 t
    \int_0^\infty\dqq\frac{\mathcal{P}(q)}{q^6}
    \ints\dnu \frac{f(\vu)}{u}
    \frac{\step(1-\eta^2-k^2/q^2)}{\sqrt{1-\eta^2-k^2/q^2}}
    \left[
    q^2\cos^2\psi - k^2\frac{\cos^2\psi-\eta^2\sin^2\psi}{1-\eta^2}
    \right]
\end{align}\begin{align}
	P_{\uvr\uvec p_1}(k) 
	&= -32\pi^3 G^2\rhob^2 k^2 t
    \int_0^\infty\dqq\frac{\mathcal{P}(q)}{q^6}
    \ints\dnu \frac{f(\vu)}{u}
    \frac{\step(1-\eta^2-k^2/q^2)}{\sqrt{1-\eta^2-k^2/q^2}}
    \frac{\eta}{\sqrt{1-\eta^2}}
    \cos\psi
\end{align}\begin{align}
	P_{\uvr\uvec p_2}(k) 
	&= -32\pi^3 G^2\rhob^2 k^2 t
    \int_0^\infty\dqq\frac{\mathcal{P}(q)}{q^6}
    \ints\dnu \frac{f(\vu)}{u}
    \frac{\step(1-\eta^2-k^2/q^2)}{\sqrt{1-\eta^2-k^2/q^2}}
    \frac{\eta}{\sqrt{1-\eta^2}}
    \sin\psi
\end{align}\begin{align}
	P_{\uvec p_1\uvec p_2}(k) 
	&= -32\pi^3 G^2\rhob^2 t
    \int_0^\infty\dqq\frac{\mathcal{P}(q)}{q^6}
    \ints\dnu \frac{f(\vu)}{u}
    \frac{\step(1-\eta^2-k^2/q^2)}{\sqrt{1-\eta^2-k^2/q^2}}
    \left[q^2-k^2\frac{1+\eta^2}{1-\eta^2}\right]\cos\psi\sin\psi.
\end{align}
Already we see that $P_{\uvr\uvr}(k)$ is given by Eq.~(\ref{Pvpara1}). Now as a simplification, let us assume that the velocity distribution $f(\vu)$ depends on direction only through $\eta=\uvu\cdot\uvr$ and not $\psi$. This is true, for instance, if anisotropy in the substructure velocity distribution arises solely due to the stellar stream's own orbital motion. In this case, we can immediately average over $\psi$, so that $\sin^2\psi$ and $\cos^2\psi$ become $1/2$ and $\sin\psi$, $\cos\psi$, and $\cos\psi\sin\psi$ become $0$. Then $P_{\uvr\uvec p_1}(k)=P_{\uvr\uvec p_2}(k)=P_{\uvec p_1\uvec p_2}(k)=0$ and $P_{\uvec p_1\uvec p_1}(k)=P_{\uvec p_2\uvec p_2}(k)$ is given by Eq.~(\ref{Pvperp1}).

\section{Velocity-injection power spectrum -- the substructure velocity integral}\label{sec:anisotropy}

The expressions for the velocity-injection power spectra, Eqs. (\ref{Pvpara1}) and~(\ref{Pvperp1}), involve integrals over the distribution of substructure velocities $\vu$ relative to the stream. In this appendix, we evaluate these integrals in the scenario where the substructure velocity distribution is isotropic \textit{in the Galactic frame}, but the stellar stream is also moving with some velocity $\vv$ that is parallel to the separation vector $\vr$ along which we are interested in correlations. That is, the Galactic-frame substructure velocity $\tilde \vu\equiv\vv+\vu$ is isotropically distributed while $\vu$ itself is not.

Defining (as in Appendix~\ref{sec:angular}) $\eta\equiv\uvu\cdot\uvr = \uvu\cdot\uvv$, we first note that
\begin{align}
    u &= \sqrt{\tilde u^2+v^2-2\tilde\eta\tilde u v}
    \ \ \text{and} \ \ 
    \eta =\frac{\tilde\eta\tilde u-v}{\sqrt{\tilde u^2+v^2-2\tilde\eta\tilde u v}},
\end{align}
where $\tilde\eta\equiv \hat{\tilde\vu}\cdot\uvv$ is the cosine of the angle between $\tilde\vu$ and $\vv$. Let $\tilde f(\tilde u)$ be the isotropic distribution of $\tilde \vu$ and recall that $f(\vu)$ is the (anisotropic) distribution of $\vu$. Since $\dnu=\diff^3\tilde\vu$ and $f(\vu)=\tilde f(\tilde u)$, the velocity integral in Eqs. (\ref{Pvpara1}) and~(\ref{Pvperp1}) becomes
\begin{align}\label{aniso_step1}
    \ints\dnu \frac{f(\vu)}{u}
    \frac{\step(1-\eta^2-x^2)}{(1-\eta^2-x^2)^{1/2}}
    &=
    \ints\diff^3\tilde\vu \tilde f(\tilde u) \int_{-1}^1\frac{\diff\tilde \eta}{2}\frac{\step[1-x^2-(\tilde\eta\tilde u-v)^2/(\tilde u^2+v^2-2\tilde\eta\tilde u v)]}{\sqrt{1-x^2-(\tilde\eta\tilde u-v)^2/(\tilde u^2+v^2-2\tilde\eta\tilde u v)}}
    \frac{1}{\sqrt{\tilde u^2+v^2-2\tilde\eta\tilde u v}}
    \\\label{aniso_step2}
    &=
    \ints\diff^3\tilde\vu \frac{\tilde f(\tilde u)}{\tilde u} \int_{-1}^1\frac{\diff\tilde \eta}{2}\frac{\step[1-\tilde\eta^2-x^2(1+y^2-2y\tilde\eta)]}{\sqrt{1-\tilde\eta^2-x^2(1+y^2-2y\tilde\eta)}},
\end{align}
where $x\equiv k/q$ and $y\equiv v/\tilde u$. In the second line, we used that $1+y^2-2y\tilde\eta>0$ for $-1<\tilde\eta<1$. The expression $1-\tilde\eta^2-x^2(1+y^2-2y\tilde\eta)$ is always negative (for $-1<\tilde\eta<1$) if either $xy>1$ or $x>1$, but otherwise it has the real factorization
\begin{align}
    1-\tilde\eta^2-x^2(1+y^2-2y\tilde\eta) = (\eta_+-\tilde\eta)(\tilde\eta-\eta_-),
    \ \ \text{where} \ \ 
    \eta_{\pm}\equiv x^2 y  \pm \sqrt{(1-x^2)(1-x^2 y^2)},
\end{align}
with $1\leq\eta_-< \eta_+\leq 1$. The velocity integral therefore becomes
\begin{align}\label{aniso_step4}
    \ints\dnu \frac{f(\vu)}{u}
    \frac{\step(1-\eta^2-x^2)}{(1-\eta^2-x^2)^{1/2}}
    &=
    \frac{1}{2}\step(1-x)
    \int\diff^3\tilde\vu \frac{\tilde f(\tilde u)}{\tilde u}
    \step(1-xy)\int_{\eta_-}^{\eta_+}\diff\tilde \eta \frac{1}{\sqrt{(\eta_+-\tilde\eta)(\tilde\eta-\eta_-)}}
    \\\label{aniso_step5} &=
    \frac{\pi}{2}\step(q-k)\int\diff^3\tilde\vu \frac{\tilde f(\tilde u)}{\tilde u}
    \step\!\left(q\tilde u-k v\right)
\end{align}
(the $\tilde\eta$ integral in Eq.~\ref{aniso_step4} evaluates to $\pi$ independently of $\eta_\pm$). Substituting this equation into Eqs. (\ref{Pvpara1}) and~(\ref{Pvperp1}) implies Eqs. (\ref{Pvpara}) and~(\ref{Pvperp}).

\section{Solving for the perturbed stream distribution function}\label{sec:f1}

In this appendix, we solve the partial differential equation Eq.~(\ref{BE1}) that describes the evolution of the perturbation $f_1$ to the stream's distribution function. First, we Fourier transform Eq.~(\ref{BE1}) over positions to obtain
\begin{equation}\label{f1eq}
    \frac{\partial f_1(k,v,t)}{\partial t} + \I k v f_1(k,v,t)
    =
    -C(k,t) \frac{\partial f_0(v,t)}{\partial v}
    + \frac{1}{2}D\frac{\partial^2 f_1(k,v,t)}{\partial v^2},
\end{equation}
which we can rewrite in the form
\begin{equation}
    \frac{\partial f_1(k,v,t)}{\partial t}
    + \left[\I k v - \frac{1}{2}D\frac{\partial^2}{\partial v^2} \right]f_1(k,v,t)
    = -C(k,t) \frac{\partial f_0(v,t)}{\partial v}.
\end{equation}
Recall that we approximate $D$ to be a constant. Using the operator-valued integrating factor $\e^{t(\I kv-\frac{1}{2}D\frac{\partial^2}{\partial v^2})}$, we can formally write the solution as
\begin{align}
    f_1(k,v,t) = 
    -\int_{0}^t\diff t^\prime 
    C(k,t^\prime)
    \e^{(t^\prime-t)(\I kv-\frac{1}{2}D\frac{\partial^2}{\partial v^2})}
    \frac{\partial f_0(v,t^\prime)}{\partial v},
\end{align}
where we assume that substructure encounters begin at $t=0$. Now define the operators $A=\I kv$ and $B=-\frac{1}{2}D\frac{\partial^2}{\partial v^2}$. The sequence of commutators of these operators terminates: $[A,B]=\I k D \partial/\partial v$, $[A,[A,B]]=k^2 D$, and all other commutators vanish. Thus, by the Zassenhaus formula,
\begin{align}
    \e^{(t^\prime-t)(A+B)} &= 
    \e^{(t^\prime-t)A}
    \e^{(t^\prime-t)B}
    \e^{-\frac{1}{2}(t^\prime-t)^2 [A,B]}
    \e^{\frac{1}{6}(t^\prime-t)^3 [A,[A,B]]}
\end{align}
\cite[e.g.][]{casas2012efficient} and hence
\begin{align}
    f_1(k,v,t) = 
    -\int_{0}^t\diff t^\prime
    C(k,t^\prime)
    \e^{\I kv(t^\prime-t)}
    \e^{-\frac{1}{2}D(t^\prime-t)\frac{\partial^2}{\partial v^2}}
    \e^{-\frac{\I}{2} k D (t^\prime-t)^2\frac{\partial}{\partial v}}
    \e^{\frac{1}{6} k^2 D (t^\prime-t)^3}
    \frac{\partial f_0(v,t^\prime)}{\partial v}.
\end{align}
But $\e^{-\frac{\I}{2} k D (t^\prime-t)^2\frac{\partial}{\partial v}}$ and $\e^{-\frac{1}{2}D(t^\prime-t)\frac{\partial^2}{\partial v^2}}$ are a shift operator and a Weierstrass transform, respectively:
\begin{align}
    \e^{-\frac{\I}{2} k D (t^\prime-t)^2\frac{\partial}{\partial v}}f(v)
    &=
    f\!\left[v-\frac{\I}{2} k D (t^\prime-t)^2\right]
    \ \ \text{and}\ \ 
    \e^{-\frac{1}{2}D(t^\prime-t)\frac{\partial^2}{\partial v^2}} f(v)
    =
    \frac{1}{\sqrt{4\pi}}\int_{-\infty}^\infty\diff z\,
    \e^{-z^2/4}
    f\!\left[v-\I\sqrt{D(t^\prime-t)/2}\,z\right]
\end{align}
(where $f$ is an arbitrary function). Consequently, defining $f_0^\prime\equiv\partial f_0/\partial v$, we find that 
\begin{align}\label{f1_gen}
    f_1(k,v,t) = 
    -\int_{0}^t\diff t^\prime
    C(k,t^\prime)
    \e^{\I kv(t^\prime-t)+\frac{1}{6} k^2 D (t^\prime-t)^3}
    \frac{1}{\sqrt{4\pi}}
    \int_{-\infty}^\infty\diff z\,
    \e^{-z^2/4}
    f_0^\prime\!\left[v-\frac{\I}{2} k D (t^\prime-t)^2-\I\sqrt{D(t^\prime-t)/2}\,z,t^\prime\right].
\end{align}
Let us finally assume that the unperturbed velocity distribution is Gaussian (Maxwellian), $f_0(v,t)=(2\pi\sigma^2)^{-1/2}\rhob_*\exp\left(-\frac{v^2}{2\sigma^2}\right)$, with velocity dispersion $\sigma^2=\sigma_0^2+D t$. Then we can evaluate the $z$ integral in Eq.~(\ref{f1_gen}) and obtain Eq.~(\ref{f1}).

\section{Unequal-time velocity-kick correlations}\label{sec:unequaltime}

In Section~\ref{sec:density}, we assumed that velocity kicks are uncorrelated if they occur at different times. We now show this explicitly. The unequal-time correlation function between velocity injections (projected along arbitrary directions $\uvec a$ and $\uvec b$) is
\begin{align}\label{corr_dv_t_}
	\langle \uvec a\sdot\Dv(0,t) \uvec b\sdot\Dv(\vr,t^\prime)\rangle
	&=
	\ints\dnq P(q)\ints\diff^3\vu f(\vu)
	\e^{-\I\vq\cdot\vec r}
	\uvec a\sdot\vV^*(\vq|\vu,t)\uvec b\sdot\vV(\vq|\vu,t^\prime)
	\\&=
	(8\pi G\rhob )^2
	\ints\dnq P(q)\ints\diff^3\vu f(\vu)
	\frac{(\uvec a\sdot\uvq)(\uvec b\sdot\uvq)}{q^2}
	\e^{-\I\vq\cdot\vec r}
	\e^{-\I\vq\cdot\vu (t-t^\prime)/2}
	\frac{\sin(\vq\cdot\vu t/2)}{\vq\cdot\vu}
	\frac{\sin(\vq\cdot\vu t^\prime/2)}{\vq\cdot\vu}
	\\&=
	(8\pi G\rhob )^2
	\ints\dnq P(q)\ints\diff^3\vu f(\vu)
	\frac{(\uvec a\sdot\uvq)(\uvec b\sdot\uvq)}{q^4 u^2}
	\e^{-\I\vq\cdot\vec r}
	f(\uvq\cdot\uvu,qut/2,qut^\prime/2),
\end{align}
where
\begin{equation}
    f(x,t_1,t_2) \equiv
    \e^{-\I x (t_1-t_2)}
	\frac{\sin(x t_1)\sin(x t_2)}{x^2}
	\to
	\pi \min(t_1,t_2)\ddelta(x)
\end{equation}
approaches a delta function when $x t_1\gg 1$ and $x t_2\gg 1$. The pre-factor $\pi \min(t_1,t_2)$ is simply the integral of $f(x,t_1,t_2)$ over $x$ (in the same limit). Thus, in the $qut\gg 1$ and $qut^\prime\gg 1$ limit,
\begin{align}\label{corr_dv_t}
	\langle \uvec a\sdot\Dv(0,t) \uvec b\sdot\Dv(\vr,t^\prime)\rangle
	&=
	32\pi^3 G^2\rhob^2 \min(t,t^\prime)
	\ints\dnq \frac{P(q)}{q^3}\ints\diff^3\vu \frac{f(\vu)}{u}
	\e^{-\I\vq\cdot\vec r}
	(\uvec a\sdot\uvq)(\uvec b\sdot\uvq)
	\ddelta(\uvq\cdot\uvu).
\end{align}
If we consider instead acceleration correlations, it follows immediately that
\begin{align}\label{corr_ddv_t}
    \left\langle
    \uvec a\sdot\frac{\diff \Dv(0,t)}{\diff t}\,
    \uvec b\sdot\frac{\diff \Dv(\vr,t^\prime)}{\diff t^\prime}
    \right\rangle 
    =
    \ddelta(t-t^\prime)
    \frac{\diff}{\diff t}
    \langle \uvec a\sdot\Dv(0,t) \uvec b\sdot\Dv(\vr,t)\rangle,
\end{align}
so velocity kicks occurring at different times are uncorrelated. Equation~(\ref{corr_ddv_t}) also clearly implies Eq.~(\ref{Ccorr}).

\section{Stream velocity power spectrum}\label{sec:pkv}

\begin{figure*}
	\centering
	\includegraphics[width=\linewidth]{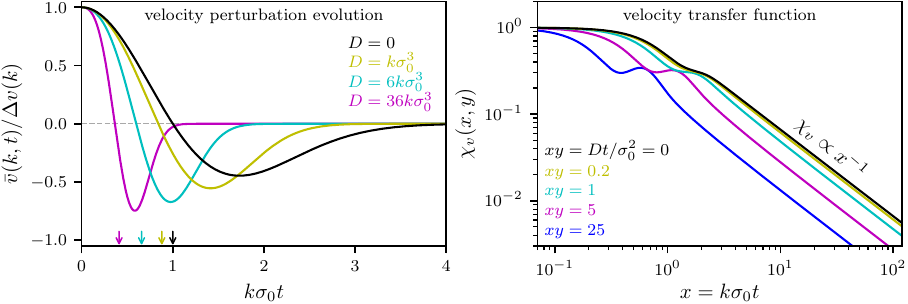}
	\caption{Similar to Fig.~\ref{fig:cutoff-d} but instead showing the behavior of stream velocity perturbations. \textit{Left-hand panel}: Time evolution of the mean velocity $\bar v$ due to a distribution $\Delta v(k)$ of velocity kicks that occur at time $t=0$. Like the density perturbation, the mean velocity is also suppressed rapidly by the system's velocity dispersion. However, there is also a sign flip when stars moving in one direction begin to drift into the positions initially occupied by stars moving in the opposite direction. This flip occurs roughly when $k s_\mathrm{rms}=1$ (small arrow; see Eq.~\ref{srms}). \textit{Right-hand panel}: The transfer function $\chi_v$ that sets the power spectrum $P_v(k,t)$ of the mean velocity within the stream (Eq.~\ref{powerv}). This function resembles the density transfer function in Fig.~\ref{fig:cutoff-d} but scales differently at large $x\equiv k\sigma_0 t$. Also, the sign flip in the velocity perturbation evolution leads to a bump-like feature near $x\sim 1$.}
	\label{fig:cutoff-v}
\end{figure*}

We noted in Section~\ref{sec:other} that a stellar stream's mean-velocity power spectrum can be computed straightforwardly in a manner similar to the density power spectrum. In particular, we can multiply Eq.~(\ref{f1}) by $v$ and integrate over velocities to obtain
\begin{align}\label{v-vdisp}
    \bar{v}(k,t) &= 
    \int_{0}^t\diff t^\prime
    \left[1-k^2(t-t^\prime)^2\left(\sigma_0^2+D\frac{t+t^\prime}{2}\right)\right]
    C(k,t^\prime)
    \e^{-k^2 (t-t^\prime)^2 [\sigma_0^2+D(t+2t^\prime)/3]/2}.
\end{align}
Evidently, the response of $\bar v$ to a distribution $\Delta v(k)$ of velocity kicks occurring at $t=0$, i.e. $C(k,t)=\Delta v(k)\ddelta(t)$, is
\begin{align}\label{v_time}
    \bar v(k,t)
    = \left[1-k^2 (\sigma_0^2 + D t/2) t^2\right]
    \e^{-k^2 (\sigma_0^2+D t/3)t^2/2} \Delta v(k)
    &&
    \text{(constant $\Delta v$)}
\end{align}
in the case where the velocity dispersion still grows as $\sigma_0^2+Dt$. We plot the resulting time evolution in the left-hand panel of Fig.~\ref{fig:cutoff-v}. Initially $\bar v = \Delta v$ before $\bar v$ becomes exponentially suppressed by the velocity dispersion. Interestingly, before this exponential suppression, $\bar v$ undergoes a sign change as particles moving in one direction drift into regions originally occupied by particles moving in the opposite direction. Roughly, this sign change occurs when the rms particle displacement due to the velocity dispersion is equal to $1/k$; that is, $ks_\mathrm{rms}=1$ (see Eq.~\ref{srms}).

For an arbitrary spectrum $P_{\Delta v,\parallel}(k,t)$ of velocity injections, the same procedure as Section~\ref{sec:densitypower} shows that the power spectrum of mean velocities is given by
\begin{align}
    P_{v,\parallel}(k,t) = 
    \int_0^t\diff t^\prime 
    \left[1-k^2(t-t^\prime)^2\left(\sigma_0^2+D\frac{t+t^\prime}{2}\right)\right]^2
    \exp\!\left[-k^2 \left(\sigma_0^2+D \frac{t+2t^\prime}{3}\right)(t-t^\prime)^2\right]
    \frac{\diff P_{\Delta v,\parallel}(k,t^\prime)}{\diff t^\prime}.
\end{align}
If we specialize to the case where $P_{\Delta v,\parallel}\propto t$, we obtain
\begin{equation}\label{powerv}
    P_{v,\parallel}(k,t) = 
    \chi_v\!\left(k\sigma_0 t,\frac{D}{k\sigma_0^3}\right)
    P_{\Delta v,\parallel}(k,t),
\end{equation}
where
\begin{align}\label{cutoffv}
    \chi_v(x,y) &\equiv 
    \frac{1}{x}\int_0^x\diff x^\prime \left[1-\left(x-x^\prime\right)^2\left(1+y\frac{x+x^\prime}{2}\right)\right]^2 \exp\!\left[-\left(x-x^\prime\right)^2\left(1+y\frac{x+2x^\prime}{3}\right)\right]
\end{align}
is a transfer function that encodes the suppression of power by the velocity dispersion $\sigma_0^2+Dt$.

We plot $\chi_v$ in the right-hand panel of Fig.~\ref{fig:cutoff-v}. In the limit that $\sigma_0^2\gg D t$, we can approximate $y=0$ and evaluate
\begin{align}
    \chi_v(x,0) &=
    \frac{3\sqrt{\pi}}{8}x^{-1}\erf(x)+\frac{1-2x^2}{4}\e^{-x^2},
\end{align}
which ranges from $\chi_v(x,0)\simeq 1$ when $x\ll 1$ to $\chi_v(x,0)\simeq (3/8)\pi^{1/2} x^{-1}$ when $x\gg 1$. It may appear notable that whereas in the velocity dispersion-dominated regime the density power spectrum is suppressed by the power $k^{-3}$ of the wavenumber, the velocity power spectrum is suppressed by only the power $k^{-1}$. However, this feature only suffices to compensate the factor of $k^2 t^2$ present in the expression for the density power spectrum, Eq.~(\ref{powerd}), that is absent from Eq.~(\ref{powerv}). That is, the density and velocity power spectra exhibit different scaling behaviours only at small, and not large, $k$. Within the growing, steady-state, and decaying phases discussed in Section~\ref{sec:densitypower},
\begin{align}\label{earlypowerv}
    P_{v,\parallel}(k,t) &= t \frac{\diff P_{\Delta v,\parallel}(k,t)}{\diff t},
    &&
    \text{(growing regime)}
    \\\label{staticpowerv}
    P_{v,\parallel}(k,t) &= 
    \frac{3\pi^{1/2}}{8}
    k^{-1}\sigma_0^{-1}
    \frac{\diff P_{\Delta v,\parallel}(k,t)}{\diff t}
    =
    \frac{3}{2}\sigma_0^2 \Ps(k,t),
    &&
    \text{(steady-state regime)}
    \\\label{decaypowerv}
    P_{v,\parallel}(k,t) &= 
    \frac{3\pi^{1/2}}{8}
    k^{-1} D^{-1/2} t^{-1/2} \frac{\diff P_{\Delta v,\parallel}(k,t)}{\diff t}
    =
    \frac{3}{2} D t \Ps(k,t).
    &&
    \text{(decaying regime)}
\end{align}

\section{Idealized simulations}\label{sec:simulation}

We performed idealized non-orbital simulations to test the analytic predictions of Sections \ref{sec:inhomogeneity} and~\ref{sec:disp}; the results of these simulations are shown in Fig.~\ref{fig:sim} and discussed in Section~\ref{sec:validation}. We now show how these simulations were executed.

For these simulations we consider $N=10^5$ stream stars arranged along a periodic line of length $L=12$~kpc. The stars have some initial velocity dispersion $\sigma_0$. These stars are subjected to encounters with a uniform distribution of extended ``subhalo'' particles. The particles are taken, for simplicity, to have a lognormal mass distribution centred about $3~\mathrm{M}_\odot$ with a standard deviation of 1~e-fold. The number density of particles is set such that the environment's average mass density is $\rhob=5\times 10^5~\mathrm{M}_\odot\,\mathrm{kpc}^{-3}$. Internally, particles are taken to be Plummer spheres with central density $100\rhob$. Finally, particles are given a Maxwellian velocity distribution with scale velocity $u_0=125~\mathrm{km}\,\mathrm{s}^{-1}$ and are further boosted by a fixed velocity $v$ along the stream's linear track (to represent the stream itself moving). These choices are purely for validation purposes and are not intended to represent realistic dark matter substructure.

``Subhalo'' particles are taken to perturb stellar velocities in the following way. Instead of treating these particles as moving objects, we directly apply the integrated velocity kick
\begin{equation}\label{sim_impulse}
	\Dv = \frac{2 G m}{u}\frac{\vec b}{b^2+r^2}
\end{equation}
to each star at the moment the particle is generated, where $m$ and $r$ are the particle's mass and Plummer scale radius, respectively, $u$ is its speed relative to the stream, and $\vec b$ is the impact parameter associated with its encounter with the subject star. Additionally, to keep the simulation finite, we only allow a star to be influenced by encounters with particles that come within the finite maximum impact parameter $\bmax=0.3$~kpc. We show in Appendix~\ref{sec:bmax} that when $qut\gg 1$, the impact of these approximations is that the environment's density power spectrum is effectively scaled by the factor $[1-J_0(\bmax q)]^2$. The power spectrum associated with our ``subhalo'' particle distribution, with this scaling applied, is plotted in Fig.~\ref{fig:sim}. Finally, since the simulation is periodic, we allow for the possibility that a single particle, depending on its velocity, may encounter a star multiple times, and we sum the velocity kicks (Eq.~\ref{sim_impulse}) from all of these encounters.

We sample encounters in the following way. We assume ``subhalo'' particles are uniformly distributed in space with number density $n=\rhob/\bar m\simeq 10^5$~kpc$^{-3}$, where $\bar m\simeq 4.9~\mathrm{M}_\odot$ is the average mass. Further, we let $f(\tilde u) \diff \tilde u$ be the isotropic Maxwellian velocity distribution described above; the added velocity $v$ along the stream's track does not affect the encounter frequency (since the stream is periodic). Within the time interval $\diff t$, the differential number of particles $N$ that cross a planar surface of area $\diff A$ is
\begin{equation}\label{cross_area}
    \diff N = n\, \diff A\, \tilde u f(\tilde u)\diff\tilde  u\, |\mu|\frac{\diff\mu}{2} \, \diff t,
\end{equation}
where $\mu$ is the cosine of the angle between the velocity vector and the normal to the plane. If we are only interested in particles crossing in one direction, $|\mu|\diff\mu/2$ integrates to $1/4$. To sample particles that come within the distance $\bmax$ of our line of stars, we use Eq.~(\ref{cross_area}) to determine the distribution of particles that cross inward through the corresponding cylinder of radius $\bmax$ and length $L$. Since the line is periodic, we do not need to sample any particles beyond the stream's linear extent.

Integrating over the other variables, we use Eq.~(\ref{cross_area}) to determine the mean time $\diff t/\diff N\simeq 8.55$~yr between encounters. We therefore execute the simulation by alternating the following two processes:
\begin{enumerate}
    \item We subject stream stars to a random ''subhalo'' particle encounter, as described above.
    \item We drift all stream stars for a time interval $\Delta t$ that is exponentially distributed with mean $\diff t/\diff N$.
\end{enumerate}
For simplicity, we allow drift only along the stream and not perpendicular to it. In this way we evolve the stream of stars for the total duration of $t=7$~Gyr, after which we measure its density and velocity power spectra and plot them in Fig.~\ref{fig:sim}.

\section{Limiting the allowable impact parameters}\label{sec:bmax}

In any numerical simulation of stream perturbations that does not resolve the full Galactic context, computational expense must be kept finite by restricting the maximum distance at which stream-substructure encounters are considered. A natural way to implement this limit is to impose a maximum impact parameter $\bmax$ for substructure encounters \citep[e.g.][]{bovy2017linear}; see also Section~\ref{sec:accuracy}. We show in this appendix how that restriction affects the heating process. To do so, we return to the derivation of the velocity-injection response function $\vV$ in Section~\ref{sec:static}. To impose a maximum impact parameter, we will find it convenient to derive $\vV$ using a different approach.

Consider the scenario under which the velocity injection $\Dv$, Eq.~(\ref{Dv_k}), was derived: a star is moving through a static density field $\rho(\vx)=\rhob[1+\delta(\vx)]$ with relative velocity $\vu$. Suppose that the star's initial position is $\vx_*=0$. Integrated over all time, an encounter with the mass element $\diff m$ imparts the star with velocity
\begin{equation}\label{impulse}
	\diff \vv = \frac{2 G \diff m}{u b^2} \vec b,
\end{equation}
where $\vec b$ is the impact parameter, which points from the star to the mass element's closest approach. Note that $\vec b$ is perpendicular to $\vu$. The velocity change integrated over all mass elements that reach closest approach within the time interval $t$ is then
\begin{align}\label{dv_imp}
	\Dv &= \frac{2G}{u} \int\diff^2 \vec b \, \step(\bmax-b) \int_{0}^{ut}\diff z\,\rho(\vx) \frac{\vec b}{b^2}
	= \int\dnx\, \delta(\vx)\vVi(\vx|\vu,t)
	= \int\dnq\, \delta(\vq)\vVi^*(\vq|\vu,t),
\end{align}
where $\vec b\equiv\vx-z\uvu$ and $z\equiv\vx\cdot\uvu$. Here, we define
\begin{equation}
    \vVi(\vx|\vu,t)\equiv\frac{2G\rhob}{u}\frac{\vec b}{b^2}\step(\bmax-b)\step(ut-z)\step(z).
\end{equation}
The Fourier transform of this expression is
\begin{align}\label{Vimp}
	\vVi(\vq|\vu,t)
	&=
	\frac{2G\rhob}{u}
	\int_{0}^{ut}\diff z\, \e^{-\I q_\parallel z} \int \diff^2 \vec b\, \e^{-\I \vq_\perp\cdot\vec b}\frac{\vec b}{b^2}\step(\bmax-b)
	=
	8\pi \I G\rhob\,\e^{\I q_\parallel ut/2}\frac{\sin(q_\parallel u t/2)}{q_\parallel u}\frac{\vq_\perp}{q_\perp^2}\left[1-J_0(\bmax q_\perp)\right],
\end{align}
where $q_\parallel \equiv \vq\cdot\uvu$, $\vq_\perp \equiv \vq - q_\parallel\uvu$, and $J_0$ is a Bessel function of the first kind. Evidently, our summation over impulsive velocity kicks instead of direct integration of the acceleration (as in Section~\ref{sec:formalism}) has replaced the $\vq/q^2$ factor in Eq.~(\ref{V}) with $\vq_\perp/q_\perp^2$. However, the $qut\gg 1$ limit enforces $\vq_\perp=\vq$, in which case the only difference between Eq.~(\ref{Vimp}) and Eq.~(\ref{V}) is the factor $\left[1-J_0(\bmax q)\right]$. Since the substructure power spectrum $P(q)$ always appears in conjunction with two factors of $\vV$, we claim that
\begin{equation}\label{Psim}
    P_\mathrm{eff}(q) = \left[1-J_0(\bmax q)\right]^2 P(q)
\end{equation}
is the effective power spectrum for the purpose of velocity injections.

We also remark that by the same calculation, a minimum impact parameter $\bmin$ may be imposed by scaling the substructure power spectrum by $[J_0(\bmin q)]^2$. For a given star mass $m_*$, a suitable choice of $\bmin$ might be $b_{90}\equiv  G m_*/u^2$, the impact parameter that leads to a 90-deg deflection of the encountering substructure mass element. At this point the assumption that substructure travels at a fixed relative velocity $\vu$ breaks down \citep[e.g.][]{binney2011galactic}. However, for the example parameters $m_*=1$~$\mathrm{M}_\odot$ and $u=120$~km\,s$^{-1}$, $b_{90}=3\times 10^{-10}$~kpc, a scale that is far below what any stellar stream can probe.

\section{Connection between local and global velocity dispersions}\label{sec:sigmalocal}

We noted in Section~\ref{sec:sigmaeff} that due to a stellar stream's global evolution, its local velocity dispersion $\sigma_0$ is smaller than its total (global) velocity dispersion $\sigma_\mathrm{tot}$ due to self-sorting. In particular, since all member stars originate at the location of the stream progenitor, faster stars tend to end up farther from the progenitor while slower stars remain closer. To estimate the connection between $\sigma_0$ and $\sigma_\mathrm{tot}$, consider the following simplified arrangement. Each star is released at position 0, time $t$ uniformly distributed in $-t_\mathrm{age}<t<0$, and velocity $v$ distributed according to some function $f(v)$. The star's position at time $0$ today is $x = -v t$. We seek the conditional distribution $f(v|x)$ of $v$ given $x$.

\begin{figure}
	\centering
	\includegraphics[width=0.5\linewidth]{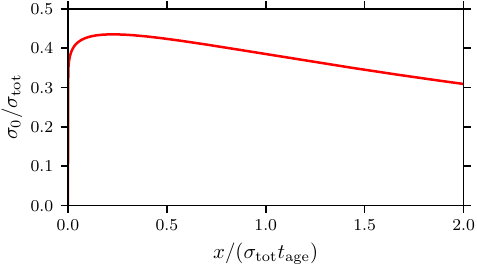}
	\caption{Local velocity dispersion $\sigma_0$ at one-dimensional position $x$ on the stream (where $x=0$ represents the progenitor's location) in units of the total velocity dispersion $\sigma_\mathrm{tot}$. We assume the total velocity distribution is normal. $\sigma_0<\sigma_\mathrm{tot}$ because particles of similar velocities tend to arrive at similar positions. In particular, we plot Eq.~(\ref{sigmalocal}), which is derived under the simplifying assumptions given in the text. $t_\mathrm{age}$ is the age of the stream.}
	\label{fig:sigma}
\end{figure}

Without loss of generality we assume $v>0$ and hence also $x>0$. For fixed $v$, the uniform distribution of $t$ implies a uniform distribution
\begin{equation}
    f(x|v) = \begin{cases} 
          1/(v t_\mathrm{age}) & \text{if}\ 0<x<v t_\mathrm{age} \\
          0 & \text{otherwise}
       \end{cases}
\end{equation}
in $x$. By Bayes' theorem,
\begin{equation}\label{vdist}
    f(v|x) = A^\prime f(x|v) f(v) = A \step(v-x/t_\mathrm{age}) f(v)/v,
\end{equation}
where $A^\prime$ and $A$ are (different) normalization constants, which depend on $x$ and $t_\mathrm{age}$ but not $v$. If we assume $f(v)$ is normal with variance $\sigma_\mathrm{tot}^2$, then by integrating moments of Eq.~(\ref{vdist}) we find that
\begin{align}\label{sigmalocal}
    \sigma_0^2(x) = S\!\left(\frac{x}{\sqrt{2}\sigma t_\mathrm{age}}\right) \sigma_\mathrm{tot}^2,
    &&
    S(y) \equiv -2\left[\Ei(-y^2)\right]^{-2}\left\{\pi\left[\erfc(y)\right]^2+\e^{-y^2}\Ei(-y^2)\right\},
\end{align}
where $\Ei(x)\equiv-\int_{-x}^\infty\diff t\, \e^{-t}/t$ is the exponential integral (which is negative for negative arguments) and $\erfc(x)\equiv (2/\sqrt{\pi})\int_x^\infty \diff t\,\e^{-t^2}$ is the complementary error function. The behaviour of $\sigma_0$ is plotted in Fig.~\ref{fig:sigma}.

For any well-populated point on the stellar stream, we expect that $x\lesssim \sigma_\mathrm{tot} t_\mathrm{age}$, in which case Fig.~\ref{fig:sigma} suggests that we can approximate $\sigma_0\simeq 0.4\sigma_\mathrm{tot}$. More precisely, the mean value of $\sigma_0/\sigma_\mathrm{tot}$ is $0.415$ for $x\in (0,\sigma_\mathrm{tot}t_\mathrm{age})$ and $0.380$ for $x\in (0,2\sigma_\mathrm{tot}t_\mathrm{age})$. In principle one should also account for the modified velocity distribution (Eq.~\ref{vdist}) in Section~\ref{sec:disp}'s treatment of stream perturbations, but for the sake of simplicity we neglect this effect and continue to assume that the unperturbed local velocity distribution is Maxwellian.


\bsp	
\label{lastpage}
\end{document}